\documentclass[aap,MSNbibl,seceqn,dvips]{arximspdf}
\usepackage{multirow,dcolumn}

\doi{10.1214/13-AAP977} 
\volume{24}
\issue{5}
\pubyear{2014}
\firstpage{2176}
\lastpage{2205}

\makeatletter
\newcommand{\rrvert}{\vert}
\newcommand{\llvert}{\vert}
\def\cal{\mathcal}
\newcolumntype{d}[1]{D{.}{.}{#1}}
\newtheorem{thmm}{Theorem}[section]
\newproclaim{dfn}[thmm]{Definition}

\newproclaim{rem}[thmm]{Remark}

\newproclaim{asm}[thmm]{Assumption}
\def\Ec{{\cal E}}

\def\reff#1{(\ref{#1})}

\makeatother

\begin{document}
\begin{frontmatter}

\title{Approximating stochastic volatility by recombinant~trees\thanksref{T1}}
\runtitle{Tree approximation of stochastic volatility}
\thankstext{T1}{Supported in part by the
European Research Council under the Grant 228053-FiRM,
by the ETH Foundation and by the Swiss Finance Institute.}

\begin{aug}
\author[a]{\fnms{Erd\.{i}n\c{c}} \snm{Aky{\i}ld{\i}r{\i}m}\corref{}\thanksref{t3}\ead[label=e1]{erdinc.akyildirim@math.ethz.ch}},
\author[c]{\fnms{Yan} \snm{Dolinsky}\thanksref{t2}\ead[label=e2]{yan.dolinsky@mail.huji.ac.il}}
\and
\author[b]{\fnms{H.~Mete} \snm{Soner}\thanksref{t1}\ead[label=e3]{mete.soner@math.ethz.ch}}

\thankstext{t3}{Also affiliated with Borsa Istanbul.
The views expressed
in this work are those
of the authors and do not necessarily reflect
those of the Borsa Istanbul or its
members.}
\thankstext{t2}{Also affiliated with Hebrew University.}
\thankstext{t1}{Also affiliated with Swiss Finance Institute, Zurich.}

\runauthor{E. Akyildirim, Y. Dolinsky and H.~M. Soner}

\affiliation{University of Zurich, ETH Zurich and ETH Zurich}

\address[a]{E. Akyildirim\\
Borsa Istanbul\\
34467 Emirgan\\
Istanbul\\
Turkey\\
\printead{e1}}

\address[b]{H. M. Soner\\
Department of Mathematics\\
ETH, Zurich\\
Ramistrasse 101\\
8092 Zurich\\
Switzerland\\
\printead{e3}}

\address[c]{Y. Dolinsky\\
Department of Statistics\\
Hebrew University\\
Mount Scopus 4416\\
Jerusalem 91905\\
Israel\\
\printead{e2}}
\end{aug}

\received{\smonth{5} \syear{2012}}
\revised{\smonth{3} \syear{2013}}

%
\begin{abstract}
A general method to construct
recombinant tree approximations
for stochastic volatility models is
developed
and applied to the Heston model
for stock price dynamics.
In this application, the
resulting approximation is a four tuple
Markov process. The first two components are
related to the stock and volatility processes and
take values in a two-dimensional
binomial tree. The other two components
of the Markov process are the increments
of random walks
with simple values in $\{-1,+1\}$.
The resulting efficient option pricing equations
are numerically implemented for general
American and European
options including the standard put and calls,
barrier, lookback and Asian-type pay-offs.
The weak and extended weak convergences
are also proved.
\end{abstract}

%


\end{frontmatter}

\section{Introduction}\label{sec:1}\setcounter{equation}{0}
\label{intro}

Contrary to many mathematical models,
the discrete counterpart
of the celebrated Black--Scholes model \cite{bs} came after its
continuous version,
and it is generally accepted that this simple binomial
approximation by Cox et al. \cite{crr}
has been instrumental in the better understanding
and the applicability of the model.
Rubinstein \cite{rub} states that
``the Black and Scholes model is widely viewed as one of
the most successful in the social sciences and
perhaps, including its binomial extension, the most
widely used formula, with embedded probabilities, in
human history.''

This widespread use and practicality
is extended by further research.
In particular,
stochastic volatility models have been
introduced to address the volatility
smiles observed in option markets
and the heavy tails and high
peaks of the
underlying asset distributions.
Hull and White \cite{hw},
Chesney and Scott \cite{cs},
Stein and Stein \cite{ss}, Heston \cite{hes} and Hagan et al. \cite{hagan}
among many others, assume a bivariate
diffusion framework in which a separate stochastic process
represents the dynamics of asset price volatility.
In all these models,
the asset price process~$S_t$\vadjust{\goodbreak} and its volatility factor
process $Y_t$ satisfy the following stochastic differential equations:
\begin{eqnarray*}
dS_t &=& S_t \bigl[ \mu \,dt + f(Y_t)
\,dW_t \bigr],
\\
dY_t &=& \mu^Y(Y_t) \,dt +
\sigma^Y(Y_t) \,dZ_t,
\end{eqnarray*}
with correlated Brownian motions $W, Z$
and different choices
for the functions $\mu^Y(y)$, $\sigma^Y(y)$ and $f(y)$.

In this paper, we construct an approach that
provides a {\emph{recombining}} tree approximation
for all stochastic volatility models of the above type.
This approximation as the Cox--Ross--Rubinstein (CRR)
model easily
constructs a discrete time financial market
that itself is arbitrage free and as such allows
for simple analysis of related complex
instruments.

For specificity,
we implement our methodology
on the Heston model.
Well known
among stochastic volatility models, it deserves
special attention because of its ability to provide closed-form
solutions for European options through Fourier transform.
This unique feature allows
for an efficient and quick calibration of the model to
European options. However, for derivative
products with early exercise features such as American options,
closed-form solutions do not exist even under the Heston model.
Hence, numerical methods such as binomial tree, finite
difference schemes or Monte Carlo simulation have to be
used to evaluate American and other exotic options under stochastic
volatility models.

In any market with a nonconstant volatility,
the CRR methodology encounters a basic difficulty.
Indeed, since the
volatility changes at each time, the nodes do not recombine on the
lattice, and this fact results in an exponential
and thus a computationally
explosive tree that cannot
be used in many realistic situations.
Nelson and Ramaswamy \cite{nr}
were the first to construct a computationally
simple binomial process which approximates
a diffusion process given in the form
\[
\label{Nelson} dY_t = \mu(Y,t)\,dt + \sigma(Y,t)\,dZ_t.
\]
They solve the node recombination problem by transforming the
process given in the above equation
into a process $X(Y,t)$
such that the instantaneous volatility of the transformed process
is constant. Hilliard and Schwartz
\cite{hs} follow this method
to develop binomial trees for continuous-time
risk-neutralized diffusion processes of
a special form.

Our main tool is to apply correlated random walks in
order to approximate diffusion processes.
A correlated random walk
is a generalized random\vadjust{\goodbreak} walk in the sense that the increments
are not identically and independently
distributed, but they only satisfy some
Markov-type of conditions. The exact definition will be given in
Section~\ref{s.formal}. These processes
naturally lie on a grid, and their Markov
structure allows for an efficient
computation of option prices.

The idea to use correlated random walks for approximating
diffusion processes goes back
to Gruber and Schweizer
\cite{GS1} and to Kusuoka \cite{K}.
In \cite{GS1}, the authors prove a convergence result
for one-dimensional diffusion processes
that satisfies stronger regularity conditions than those
that appear in stochastic volatility models.
In \cite{K}, Kusuoka uses (also in one dimension)
an original technique to modify random walks in order
to get a diffusion in the limit.
Again the regularity conditions that
he assumes are stronger
than those that are required in stochastic volatility models.

Our approach is also similar to that of Kusuoka
and modifies the correlated
random walks on a multi-dimensional
binomial tree by adding a predictable process
times $\sqrt{h}$ where $h$ is the size of the time
step. We then use this freedom
to choose the predicable process
together with an appropriate choice
of the conditional probabilities
to construct a Markov process
that weakly converges to the stochastic volatility model.
This construction is explained in
Section~\ref{s.formal}. The weak convergence of
our approximation is given in Section~\ref{s.proof}. Then the approximating
martingale measures are constructed so that
the modified tree under these measures
asymptotically matches the first two conditional
moments. This fact allows for a
straightforward convergence proof.
We also note that this approach was
successively used by the last two authors
\cite{DS} to prove convergence
of a market with trading costs.

Our extensive numerical
experimentation is reported in our
final section. In general, weak convergence does not provide
any error estimation. However, binomial-type approximations
of diffusion models have a convergence rate of $(\Delta t)^{{1}/{2}}$ which
we accept it to be true.
We leave the detailed description
of the computational studies
to that section and here simply
state that our algorithm works
efficiently compared
to all existing methods for the Heston model.

We emphasize that our tool can also be applied
for a general type of stochastic volatility models (see Remark \ref
{hestonextension}).
There is also GARCH approach to stochastic volatility
models that we refer the reader to
Duan \cite{d95,d96a,d96b,d97}, Nelson \cite{nelson},
Ritchken and Trevor \cite{rt}
and the references therein.

Clearly, there are several other successful
computational
approaches to stochastic models,
including the ones based on partial differential
equations, semi-analytic
methods and Monte Carlo simulations.
Here we do not survey all these
results but compare our numerical results
with the appropriate ones in the section
that outlines our numerical experimentations.

In the literature, tree-based methods have also been
considered.
Beliaeva and Nawalkha \cite{bn}
authored the most recent of these studies; see \cite{bn}
and the references therein.
However, our
approach differs from these earlier
studies in two fundamental ways.
First, our approximation is recombinant
by construction, while in the previous studies
recombination is achieved through truncation.
Also, our tree is arbitrage free, and we provide
a proof of convergence.
%

%

\section{The Heston model}\label{sec:2}
\setcounter{equation}{0}

Consider the Heston model,
\begin{eqnarray*}
d S_t &=& S_t(r \,dt+\sqrt\nu_t
\,dW_t),
\\
d\nu_t &=& \kappa(\theta-\nu_t)\,dt+ \eta\sqrt
\nu_t \,d \widetilde{W}_t,
\end{eqnarray*}
with initial conditions $S_0,\nu_0>0$,
given positive parameters $r,\kappa,\theta,\eta$
and two Brownian motions
$W, \widetilde{W}$
with a constant correlation $\rho\in(-1,1)$.
The constant $r>0$
is the interest rate
and $S$ is the stock price process.
As it is standard, we also assume that
\[
2\kappa\theta> \eta^2.
\]
Then, the Heston equation has a unique
\emph{positive} solution
in $\mathbb{R}^2_{+}$; see, for instance,~\cite{CIR}.

The main goal of this paper
is to construct a discrete approximation
of this model.
For this purpose,
it is more convenient to
work with a transformed
system of affine equations
driven by independent
Brownian motions.
Therefore, we set
\[
x_t:= \ln S_t,\qquad y_t:= \frac{\nu_t}{\eta}-
\rho x_t,
\]
so that
%
\begin{eqnarray}
\label{transformedSDE1} dx_t &=& \mu_x(x_t,y_t)\,dt+
\sqrt\eta\sigma(x_t,y_t)\,dW_t,
\nonumber
\\[-8pt]
\\[-8pt]
\nonumber
dy_t &=& \mu_y(x_t,y_t)\,dt+
\sqrt{\eta\bigl(1-\rho^2\bigr)} \sigma(x_t,y_t)\,dB_t,
\end{eqnarray}
where
\begin{eqnarray*}
\mu_x(x,y)&:=& r-\frac{1}2 \eta(y+\rho x),\qquad
\mu_y(x,y):= \frac
{\kappa
\theta}{\eta}-\rho r + \frac{1}2 (\rho\eta-2
\kappa ) (y+\rho x),
\\
B_t&:=& \frac{W_t-\rho\widetilde{W}_t}{\sqrt{1-\rho^2}},\qquad \sigma(x,y) := \sqrt{(y+\rho
x)^{+}},
\end{eqnarray*}
and $z^{+}=\max(0,z)$. One may directly verify that
$B$ is also a standard Brownian motion
independent of $W$.

\section{Derivation of the approximation}
\label{s.formal}
We fix a time horizon, or equivalently a maturity, $T>0$
and a time discretization
\[
h:= \frac{T}{n},\vadjust{\goodbreak}
\]
with a large integer $n$.
We then use
two-dimensional correlated random walks
to approximate
the diffusion processes given by \reff{transformedSDE1}.
Indeed, consider the random walks
${\{X^{(n)}_k,Y^{(n)}_k\}}_{k=0}^n$
of the form
%
\begin{eqnarray}
\label{Xtree} X^{(n)}_k&:=&x_0+ \sqrt{h\eta}
\sum_{i=1}^k \xi^X_i,
\\
\label{Ytree} Y^{(n)}_k&:=&y_0+ \sqrt{h \eta
\bigl(1-\rho^2\bigr)}\ \sum_{i=1}^k
\xi^Y_i,
\end{eqnarray}
where $x_0:= \ln(s_0)$,
$y_0:= (\nu_0/\eta) - \rho x_0$
and
$(\xi^X,\xi^Y)$'s
are random variables with values in $\{-1,1\}$.
In the sequel, we always use the initial data
\[
\xi^X_0=\xi^Y_0=0.
\]

We construct a probabilistic structure so that
the four tuple
$(X^{(n)}_k,Y^{(n)}_k,\break \xi^X_k,\xi^Y_k)$
forms a Markov chain weakly approximating
the solution of \reff{transformedSDE1}.
To achieve this we also need to introduce a
modification of this discrete
Markov chain. Indeed, for given
\emph{predictable} processes
$\hat{\alpha}, \hat{\beta}$,
we introduce
%
\begin{eqnarray}
\label{Xhattree} \hat{X}^{(n)}_k &:=& X^{(n)}_k+
\sqrt{h \eta} \hat{\alpha}_k \xi^X_k,
\\
\label{Yhattree} \hat{Y}^{(n)}_k &:=& Y^{(n)}_k+
\sqrt{h \eta\bigl(1-\rho^2\bigr)} \hat{\beta}_k
\xi^Y_k, \qquad k=1,\ldots,n.
\end{eqnarray}
Clearly, the convergence of
$(X, Y)$ is equivalent to that of
$(\hat X, \hat Y)$ as
\[
\bigl\|\hat{X}^{(n)}-X^{(n)}\bigr\|=O(\sqrt h), \qquad\bigl\|\hat{Y}^{(n)}-Y^{(n)}
\bigr\|=O(\sqrt h),
\]
where for any exponent $k$,
we use the standard notation $O(h^k)$
to denote a generic random variable of the order
$ h^k$ and $o(h^k)$ denotes
a random variable that converges to zero after divided
by $h^k$.

Our goal is to construct a sequence of
probability measures $\mathbb{P}^{(n)}$ and
stochastic processes $\hat\alpha^{(n)},\hat\beta^{(n)}$ such that
\[
\bigl\{\bigl(\hat{X}^{(n)}_{[nt/T]},\hat{Y}^{(n)}_{[nt/T]}
\bigr)\bigr\}_{t=0}^T \Rightarrow{\bigl\{(x_t,y_t)
\bigr\}}_{t=0}^T,
\]
where $\Rightarrow$ denotes weak convergence.
We provide the definitions in
the next section.

In view of the martingale
convergence Theorem
7.4.1 in \cite{EK}, to establish this convergence,
it is essentially sufficient
to match the first and the second conditional moments.
Indeed, for a positive integer $k$, set
\[
\mathcal{F}_k=\sigma\bigl\{\xi^X_1,\ldots,
\xi^X_k,\xi^Y_1,\ldots,
\xi^Y_k\bigr\},
\]
and let $\mathbb{E}^{(n)}_k[ \cdot]$
be the conditional expectation
$\mathbb{E}^{(n)}[ \cdot|\mathcal{F}_k]$
with respect to the probability measure $\mathbb{P}^{(n)}$.
Then the moment matching conditions\vadjust{\goodbreak} are the following equations:
%
\begin{eqnarray}
\label{1x} \mathbb{E}^{(n)}_{k-1}\bigl[\hat
X^{(n)}_k-\hat X^{(n)}_{k-1}\bigr] &=&
\mu_x\bigl(X^{(n)}_{k-1},Y^{(n)}_{k-1}
\bigr)h + o(h),
\\
\label{1y} \mathbb{E}^{(n)}_{k-1}\bigl[\hat
Y^{(n)}_k-\hat Y^{(n)}_{k-1}\bigr] &=&
\mu_y\bigl(X^{(n)}_{k-1},Y^{(n)}_{k-1}
\bigr)h + o(h),
\\
\label{2x} \mathbb{E}^{(n)}_{k-1} \bigl[\bigl(\hat
X^{(n)}_k-\hat X^{(n)}_{k-1}
\bigr)^2 \bigr] &=& \eta\sigma^2\bigl(X^{(n)}_{k-1},Y^{(n)}_{k-1}
\bigr)h + o(h),
\\
\label{2y}\mathbb{E}^{(n)}_{k-1} \bigl[\bigl(\hat
Y^{(n)}_k-\hat Y^{(n)}_{k-1}
\bigr)^2 \bigr] &=& \eta\bigl(1-\rho^2\bigr)
\sigma^2\bigl(X^{(n)}_{k-1},Y^{(n)}_{k-1}
\bigr)h + o(h).
\end{eqnarray}

We also need conditions on the covariances.
However, since $W$ and $B$ in \reff{transformedSDE1}
are independent, this condition is simply reduced to
the requirement that
$\xi^X_k$ and $\xi^Y_k$ are conditionally independent
given $\mathcal{F}_{k-1}$.

Observe that we need to solve four
equations, and the number of
unknowns or parameters to choose are
four as well; the corrections
$\hat\alpha, \hat\beta$ and two
probabilities,
%
\begin{equation}
\label{e.prob} p_k:= \mathbb{P}^{(n)}_{k-1}\bigl(
\xi^X_k=1\bigr), \qquad q_k:= \mathbb{P}^{(n)}_{k-1}
\bigl(\xi^Y_k=1\bigr).
\end{equation}

This construction would provide a financial
market which is asymptotically arbitrage free.
However, a slight modification of the above
procedure would also ensure that
each discrete market
itself is free of arbitrage. In our model,
the discrete stochastic process
\[
{\bigl\{\exp(-rkh)\exp\bigl(\hat{X}^{(n)}_k\bigr)\bigr
\}}_{k=0}^n,
\]
is the
approximation of the discounted
price process.
Hence, we replace the
first order condition \reff{1x}
by requiring that
above process is a martingale, that is,
%
\begin{equation}
\label{1xm} \mathbb{E}^{(n)}_{k-1} \bigl[\exp(-rh)\exp
\bigl(\hat X^{(n)}_k\bigr)-\exp\bigl(\hat
X^{(n)}_{k-1}\bigr)\bigr] =0.
\end{equation}
In fact, \reff{1x} and \reff{1xm} are asymptotically
equivalent and both would be sufficient
to prove convergence. However, in our numerical
experimentation we observe that this modification
is substantially better than the nonmodified version.
We continue by constructing
$\mathbb{P}^{(n)}$ and
$\hat\alpha^{(n)},\hat\beta^{(n)}$
satisfying equations
\reff{1xm} and \reff{1y}--\reff{2y}.
Indeed, by \reff{1xm}
we directly calculate that
\[
(1+\hat\alpha_k)\mathbb{E}^{(n)}_{k-1}\bigl[
\xi^X_k\bigr] -\hat\alpha_{k-1}
\xi^X_{k-1}= o(h).
\]
Hence
\[
\bigl(1+\hat\alpha^{(n)}_k\bigr) (\hat\alpha_{k-1})
\mathbb{E}^{(n)}_{k-1}\bigl[\xi^X_k
\bigr] \xi^X_{k-1} = (\hat\alpha_{k-1})^2
+ o(h).
\]
We use this
and calculate that
\[
\mathbb{E}^{(n)}_{k-1} \bigl(\bigl(\hat X^{(n)}_k-
\hat X^{(n)}_{k-1}\bigr)^2 \bigr) = \eta h \bigl((1+
\hat\alpha_k)^2- \bigl(\hat\alpha^{(n)}_{k-1}
\bigr)^2 +o(h) \bigr)+o(h).
\]
We expect that the difference $\hat\alpha_k - \hat\alpha_{k-1}$ to be
of order $h$.
Hence, the above expression simplifies to
\[
\mathbb{E}^{(n)}_{k-1} \bigl(\bigl(\hat X^{(n)}_k-
\hat X^{(n)}_{k-1}\bigr)^2 \bigr) = \eta h (1+2 \hat
\alpha_k )+o(h).
\]
We now compare the above equation with \reff{2x}
to conclude that
\[
1+2 \hat\alpha_k =\sigma^2\bigl(X^{(n)}_{k-1},Y^{(n)}_{k-1}
\bigr) +o(h).
\]
Using \reff{1y} and \reff{2y},
we obtain the same equation for
$\hat\beta$. Hence, we conclude that
\[
\hat\alpha_k=\hat\beta_k= \frac{\sigma^2(X^{(n)}_{k-1},Y^{(n)}_{k-1})-1}{2}+o(h).
\]
We use the above identity and
the freedom on the order $o(h)$ to define
the processes $\hat\alpha, \hat\beta$
below.
The below definition contains
a certain truncation
that is within the $o(h)$ margin.
Although this correction
is asymptotically small,
it allows us
to obtain several bounds in the convergence proof
and also enables
to construct transition probabilities
that always remain in the unit interval; see \reff{ksiXprob}, below.
So we now define
%
\begin{equation}
\label{alphabetadefinition} \hat\alpha_k:=\hat\beta_k:=
\frac{\max \{A_n,
\sigma^2(X^{(n)}_{k-1},Y^{(n)}_{k-1}) \}-1}{2},\qquad 1\leq k\leq n,
\end{equation}
where
\[
A_n= \biggl(\frac{\kappa\theta}{\eta}+ |\rho| r \biggr)\sqrt
\frac{h}{\eta(1-\rho^2)},
\]
and we set
\[
\hat\alpha^{(n)}_0=\hat\beta^{(n)}_0=0.
\]
To reiterate once again,
the function $A_n$ is chosen
to ensure that the probabilities
that are defined in \reff{ksiXprob}, below,
remain in the unit interval.
Although, this is clearly crucial
for our analysis,
in our numerical implementation
we do not use this truncation
and instead modify
\reff{ksiXprob} to ensure that
these are true probabilities.

The above construction together with
the conditional
independence of the increments
ensure the second moment
matching. We now use
the first order conditions~\reff{1xm} and \reff{1y}
to construct the transition probabilities.
Indeed, recall that by~\reff{e.prob},
\[
p_k:= \mathbb{P}^{(n)}_{k-1}\bigl(
\xi^X_k=1\bigr),
\]
and rewrite \reff{1xm} as
\begin{eqnarray*}
&& p_k \exp \bigl( \sqrt{h\eta} \bigl[ (1+ \hat\alpha_k)
- \hat\alpha_{k-1}\xi ^X_{k-1}\bigr] \bigr)
\\
&&\qquad{} +(1-p_k )\exp \bigl( - \sqrt{h\eta} \bigl[ (1+ \hat
\alpha_k) + \hat\alpha_{k-1}\xi ^X_{k-1}
\bigr] \bigr) = \exp(rh).
\end{eqnarray*}
This implies that $p_k$ must be given by
%
\begin{equation}
\label{ksiXprob} p_k = \frac{\exp(r h+\sqrt{\eta h} \hat\alpha_{k-1}\xi^X_{k-1})-
\exp(- \sqrt{\eta h } (1+\hat\alpha_{k}))} {
\exp( \sqrt{\eta h } (1+\hat\alpha_{k}) )-
\exp(- \sqrt{\eta h } (1+\hat\alpha_{k}) )}.
\end{equation}
In view of the truncation introduced in \reff{alphabetadefinition},
$p_k \in[0,1]$ for all large $n$.

We now recall that
\[
q_k:= \mathbb{P}^{(n)}_{k-1}\bigl(
\xi^Y_k=1\bigr),
\]
and use \reff{1y} to arrive at
\[
q_k= \frac{1}{2}+ \frac{\hat\alpha_{k-1}}{2 (1+\hat\alpha_{k})}\xi^Y_{k-1}+
\frac{\sqrt{h} \mu_y(X^{(n)}_{k-1},Y^{(n)}_{k-1})} {
2 \sqrt{\eta(1-\rho^2)} (1+\hat\alpha_{k})}.
\]
Since $q_k$ must take values in the unit
interval, we modify it in the following way:
%
\begin{equation}
\label{ksiYprob}\quad  q_k= \biggl(\min \biggl\{1,\frac{1}{2}+
\frac{\hat\alpha_{k-1}}{2 (1+\hat\alpha_{k})}\xi^Y_{k-1}+ \frac{\sqrt{h} \mu_y(X^{(n)}_{k-1},Y^{(n)}_{k-1})} {
2 \sqrt{\eta(1-\rho^2)}(1+\hat\alpha_{k})} \biggr\}
\biggr)^{+}.
\end{equation}
Set
\[
\Xi_{k} := \bigl(X^{(n)}_{k},Y^{(n)}_{k},
\xi^X_{k},\xi^Y_{k}\bigr).
\]
Then, we claim that $\Xi$ is a Markov process.
Indeed, recall that the independence of the Brownian motions in
\reff{transformedSDE1} implies the conditional independence
of the increments $\xi^X$ and $\xi^Y$.
Hence
%
\begin{equation}
\label{independence} \mathbb{P}^{(n)} \bigl(\xi^X_{k}=a,
\xi^Y_k=b | \Xi_{k-1} \bigr)=
\mathbb{P}^{(n)}_{k-1}\bigl(\xi^X_k=a
\bigr) \mathbb{P}^{(n)}_{k-1}\bigl(\xi^Y_k=b
\bigr).
\end{equation}
Moreover, in view of
\reff{Xtree} and \reff{Ytree}, the set
\[
\bigl\{ X_k^{(n)} =X_{k-1}^{(n)} +c,
Y_k^{(n)} =Y_{k-1}^{(n)} +d, \xi
^X_{k}=a, \xi^Y_k=b \bigr\}
\]
is empty unless $c=a\eta h$ and $d=b\eta h\sqrt{1-\rho^2}$,
and in this case it is
equal to $\{ \xi^X_{k}=a, \xi^Y_k=b \}$.
Therefore, the transition probabilities
of the process $\Xi$ are determined by
\[
\mathbb{P}^{(n)} \bigl(\xi^X_{k}=1,
\xi^Y_k=1 | \Xi_{k-1} \bigr) = p_k
q_k.
\]
Moreover, there is a simple transformation between
$\Xi_k$ and
\[
\hat\Xi_k:= \bigl(\hat X^{(n)}_{k}, \hat
Y^{(n)}_{k},\xi^X_{k},
\xi^Y_{k}\bigr).
\]
Hence, one may consider the process $\hat\Xi$
as the basic approximating Markov process.

\section{Main convergence result}
\label{s.proof}

In this section, we first briefly recall the concept of weak convergence
of probability measures and then state our main
convergence result. For more information on weak
convergence, we refer the reader to the books of Billingsley \cite{B}
and Ethier and Kurtz \cite{EK}.

For any \textit{c\`{{a}}dl\`{{a}}g}
stochastic process ${\{Z(t)\}}_{t=0}^T$ with values in
some Euclidean space
$\mathbb{R}^d$, let $\mathbb{P}^Z$
be the distribution of $Z$ on the canonical
space $\mathbb{D}([0,T];\mathbb{R}^d)$
equipped with the Skorohod
topology (for details see \cite{B}),
that is, for any Borel set $D\subset
\mathbb{D}([0,T];\mathbb{R}^d)$, $\mathbb{P}^Z(D)
=\mathbb{P}\{Z\in{D}\}$.
For a sequence of $\mathbb{R}^d$-valued,
stochastic processes
$Z^{(n)}$ we
use the notation $Z^{(n)}\Rightarrow{Z}$ to
indicate that the
probability measures $\mathbb{P}^{Z^{(n)}}$, converge
vaguely to $\mathbb{P}^Z$
on the space $\mathbb{D}([0,T];\mathbb{R}^d)$.

We are now ready to state the main convergence
theorem which is the main
theoretical foundation of
our numerical scheme. It will be proved in
Section~\ref{sec:5}.

\begin{thmm}\label{thmm2.1}
For any $n\in\mathbb{N}$,
let $\mathbb{P}^{(n)}$
be the probability measure defined
by \reff{independence}.
Consider the stochastic processes
$\{X^{(n)}_{[nt/T]}\}_{t=0}^T$,
$\{\hat X^{(n)}_{[nt/T]}\}_{t=0}^T$
and $\{Y^{(n)}_{[nt/T]}\}_{t=0}^T$
under $\mathbb{P}^{(n)}$.
Let $(x,y)$ be the unique solution
of \reff{transformedSDE1}.
Then
%
\begin{equation}
\label{2.11} \bigl\{\bigl(X^{(n)}_{[nt/T]},Y^{(n)}_{[nt/T]}
\bigr)\bigr\}_{t=0}^T \Rightarrow{\bigl\{(x_t,y_t)
\bigr\}}_{t=0}^T
\end{equation}
and
%
\begin{equation}
\label{2.12} \bigl\{\bigl(\hat X^{(n)}_{[nt/T]},Y^{(n)}_{[nt/T]}
\bigr)\bigr\}_{t=0}^T \Rightarrow{\bigl\{(x_t,y_t)
\bigr\}}_{t=0}^T
\end{equation}
on the space $\mathbb{D}([0,T])\times\mathbb{D}([0,T])$.
\end{thmm}
%
\begin{rem}
\label{hestonextension}
For the Heston model, one applies a transformation that
decorrelates
the Brownian motions. However, this
decorrelation is not necessary and used only
to simplify the procedure.
Indeed, consider a general two-dimensional diffusion
\begin{eqnarray*}
dx_t &=& \mu_x(x_t,y_t)\,dt+
\sigma_x(x_t,y_t)\,dW_t,
\\
dy_t &=& \mu_y(x_t,y_t)\,dt+
\sigma_y(x_t,y_t)\,d\widetilde{W}_t,
\nonumber
\end{eqnarray*}
where $W, \widetilde{W}$
are two-standard Brownian motions with a
correlation
$\rho$. Introduce the
two-dimensional correlated random walk
${\{X^{(n)}_k,Y^{(n)}_k\}}_{k=0}^n$
by
\begin{eqnarray*}
X^{(n)}_k&:=&x_0+ \sqrt{h} \sum
_{i=1}^k \xi^X_i,
\\
Y^{(n)}_k&:=&y_0+ \sqrt{h} \sum
_{i=1}^k \xi^Y_i.
\end{eqnarray*}
As before, we consider a small
modification of the correlated random walks
\begin{eqnarray*}
\hat{X}^{(n)}_k &:=& X^{(n)}_k+
\sqrt{h } \hat{\alpha}_k \xi^X_k,
\\
\hat{Y}^{(n)}_k &:=& Y^{(n)}_k+
\sqrt{h }\hat\beta_k \xi^Y_k,\qquad k=1,\ldots,n.
\end{eqnarray*}
In this case, the moment matching conditions
are the following equations:
\begin{eqnarray*}
\mathbb{E}^{(n)}_{k-1}\bigl[\hat X^{(n)}_k-
\hat X^{(n)}_{k-1}\bigr] &=& \mu_x
\bigl(X^{(n)}_{k-1},Y^{(n)}_{k-1}\bigr)h +
o(h),
\\
\mathbb{E}^{(n)}_{k-1}\bigl[\hat Y^{(n)}_k-
\hat Y^{(n)}_{k-1}\bigr] &=& \mu_y
\bigl(X^{(n)}_{k-1},Y^{(n)}_{k-1}\bigr)h +
o(h),
\\
\mathbb{E}^{(n)}_{k-1} \bigl[\bigl(\hat X^{(n)}_k-
\hat X^{(n)}_{k-1}\bigr)^2 \bigr] &=&
\sigma^2_x\bigl(X^{(n)}_{k-1},Y^{(n)}_{k-1}
\bigr)h + o(h),
\\
\mathbb{E}^{(n)}_{k-1} \bigl[\bigl(\hat Y^{(n)}_k-
\hat Y^{(n)}_{k-1}\bigr)^2 \bigr] &=&
\sigma^2_y\bigl(X^{(n)}_{k-1},Y^{(n)}_{k-1}
\bigr)h + o(h),
\\
\mathbb{E}^{(n)}_{k-1} \bigl[\bigl(\hat X^{(n)}_k-
\hat X^{(n)}_{k-1}\bigr) \bigl(\hat Y^{(n)}_k-
\hat Y^{(n)}_{k-1}\bigr) \bigr]& =& \sigma_x
\bigl(X^{(n)}_{k-1},Y^{(n)}_{k-1}\bigr)\sigma
_y\bigl(X^{(n)}_{k-1},Y^{(n)}_{k-1}
\bigr) \rho h + o(h).
\end{eqnarray*}
We solve these equations as in the Heston case
and obtain that
\[
\hat\alpha_k=\frac{\sigma^2_x(X^{(n)}_{k-1},Y^{(n)}_{k-1})-1}{2},\qquad
 \hat\beta_k=
\frac{\sigma^2_y(X^{(n)}_{k-1},Y^{(n)}_{k-1})-1}{2}.
\]
The transition probabilities are also given by
\begin{eqnarray*}
\mathbb{P}^{(n)}_{k-1}\bigl(\xi^X_k=1,
\xi^Y_k=1\bigr) & = & \frac{1}4+
\frac{\hat\alpha_{k-1}\xi^X_{k-1}+\mu_x \sqrt {h}}{4(1+\hat\alpha_k)}+ \frac{\hat\beta_{k-1}\xi^Y_{k-1}+\mu_y \sqrt{h}}{4(1+\hat\beta
_k)}
\\
&&{}+ \frac{\rho\sigma_x \sigma_y+\hat\alpha_{k-1}\hat\beta_{k-1}\xi
^X_{k-1}\xi^Y_{k-1}} {
4(1+\hat\alpha_k)(1+\hat\beta_k)},\\
\mathbb{P}^{(n)}_{k-1}\bigl(\xi^X_k=1,
\xi^Y_k=-1\bigr) &=& \frac{1}4 +
\frac{\hat\alpha_{k-1}\xi^X_{k-1}+\mu_x \sqrt{h}}{4(1+\hat\alpha_k)} - \frac{\hat\beta_{k-1}\xi^Y_{k-1}+\mu_y \sqrt{h}}{4(1+\hat\beta
_k)}
\\
&&{}-\frac{\rho\sigma_x \sigma_y + \hat\alpha_{k-1}\hat\beta
_{k-1}\xi
^X_{k-1}\xi^Y_{k-1}} {
4(1+\hat\alpha_k)(1+\hat\beta_k)},\\
\mathbb{P}^{(n)}_{k-1}\bigl(\xi^X_k=-1,
\xi^Y_k=1\bigr) &=& \frac{1}4-\frac{\hat\alpha_{k-1}\xi^X_{k-1}+\mu_x \sqrt {h}}{4(1+\hat\alpha_k)}
+ \frac{\hat\beta_{k-1}\xi^Y_{k-1}+\mu_y \sqrt{h}}{4(1+\hat\beta
_k)}
\\
&&{}-\frac{\rho\sigma_x \sigma_y + \hat\alpha_{k-1}\hat\beta
_{k-1}\xi
^X_{k-1}\xi^Y_{k-1}} {
4(1+\hat\alpha_k)(1+\hat\beta_k)},\\
\mathbb{P}^{(n)}_{k-1}\bigl(\xi^X_k=-1,
\xi^Y_k=-1\bigr)&=& \frac{1}4 -\frac{\hat\alpha_{k-1}\xi^X_{k-1}+\mu_x\sqrt {h}}{4(1+\hat\alpha_k)}
- \frac{\hat\beta_{k-1}\xi^Y_{k-1}+\mu_y \sqrt{h}}{4(1+\hat\beta
_k)}
\\
&&{}+\frac{\rho\sigma_x \sigma_y + \hat\alpha_{k-1}\hat\beta
_{k-1}\xi
^X_{k-1}\xi^Y_{k-1}} {
4(1+\hat\alpha_k)(1+\hat\beta_k)},
\end{eqnarray*}
where in the above formulas, functions $\mu_x, \mu_y, \sigma_x,
\sigma
_y$ are all evaluated at $(X^{(n)}_{k-1},Y^{(n)}_{k-1})$.
However, the above terms do not necessarily lie in the interval
$[0,1]$. In that case, we apply a truncation
of the form $\min(1,\max(0,\cdot))$.
\end{rem}

\begin{rem}\label{rem2.1}
We emphasize that
our approximation method using
correlated random walks
and the above convergence result
can easily be extended to more
general multidimensional
diffusions. The key idea is
the introduction of
$\hat X$-type processes
which differ from the original random walk $X$ only by
a predictable process $\hat\alpha$ times
the increment $\xi^X$. We then use this
freedom (viz., the function $\hat\alpha$)
to construct transition
probabilities that match the first and the
second conditional moments
of the original diffusion.
The approximating process has
essentially the same dimension
as the original diffusion process.
However, we need to augment
the state space by adding the increments
like $\xi^X$. But these increments
take values in the discrete set $\{-1,+1\}$
so do not increase the complexity of the
approximation.
\end{rem}

Our next remark is toward
American options.

\begin{rem}\label{rem2.2}
In general, the usual weak
convergence is not sufficient for the convergence
of American options prices.
Indeed, the
latter also requires the ``good'' behavior of the filtrations.
In his
unpublished manuscript (see \cite{A}, Sections~15--16),
David Aldous introduced the
concept of extended weak convergence
to address this problem.
Briefly his definition is as follows.
A sequence
$Z^{(n)}\dvtx \Omega_n\rightarrow\mathbb{D}([0,T];\mathbb{R}^d)$,
\emph{extended weak converges} to a
stochastic process
$Z\dvtx \Omega\rightarrow\mathbb{D}([0,T];\mathbb{R}^d)$,
if for any $k$
and continuous bounded functions
$\psi_1,\ldots,\psi_k\in{C(\mathbb{D}([0,T];\mathbb{R}^d))}$,
\[
\bigl(Z^{(n)},Z^{n,1},\ldots,Z^{n,k}\bigr)\Rightarrow
\bigl(Z,Z^{(1)},\ldots,Z^{(k)}\bigr) \qquad\mbox{in } \mathbb{D}
\bigl([0,T];\mathbb{R}^{d+k}\bigr),
\]
where for any $t\leq{T}$, $1\leq i\leq{k}$ and $n\in\mathbb{N}$,
\[
Z^{n,i}_t=\hat{E}^{(n)}\bigl(\psi_i
\bigl(Z^{(n)}\bigr) | \mathcal{F}^{Z^{(n)}}_t\bigr),\qquad
Z^{(i)}_t=\hat{E}\bigl(\psi_i(Z) |
\mathcal{F}^Z_t\bigr),
\]
$\hat{E}^{(n)}$ denotes the expectation
on the probability space on which $Z^{(n)}$ is defined
and
$\hat{E}$ denotes the expectation on the probability space on
which $Z$ is defined.
In the formulas above $\mathcal{F}^{Z^{(n)}}$ and
$\mathcal{F}^{Z}$ are the filtrations which are generated
by $Z^{(n)}$ and $Z$, respectively.
The notion of extended weak convergence
provides (in addition to the standard weak convergence
of stochastic processes)
convergence of filtrations. In particular,
Aldous proved (see \cite{A}, Section~17)
that under uniform integrability of the payoffs,
extended weak convergence implies convergence of
optimal stopping values.
However, it is known that when the proof of
weak convergence relies on martingale techniques
(like our proof), then the standard weak
convergence implies extended weak convergence.
For details, we refer the reader to
\cite{A}, Section~21.
\end{rem}


\section{Discrete pricing equations}
\label{sec:3}\setcounter{equation}{0}
In this section, we
apply the approximation developed in Section~\ref{s.formal}
to price American put and lookback options.

\subsection{American put}
Consider an American put option with a
strike price $K$. We are interested in
approximating its value given by
\[
V=\sup_{\tau\in\mathcal{T}_{[0,T]}} \mathbb{E} \bigl(e^{-r\tau}(K-S_{\tau})^{+}
\bigr),
\]
where $\mathcal{T}_{[0,T]}$ is the set of all stopping times with
respect to the filtration generated by $S$,
with values in the set $[0,T]$.
We approximate the
discounted stock price
by the discrete time martingales
\[
{\bigl\{e^{- r k h } e^{\hat{X}^{(n)}_k}\bigr\}}_{k=0}^n,\qquad
n\in\mathbb{N},
\]
constructed in Section~\ref{s.formal}.
For any $n\in\mathbb{N}$, let $\mathcal{T}_n$
be the set of all stopping times with respect to the filtration
$\mathcal{F}_k$ (again constructed in Section~\ref{s.formal}),
with values in the set $\{0,1,\ldots,n\}$.
Define
\[
V^{(n)}:=\max_{\tau\in\mathcal{T}_{n}} \mathbb{E}^{(n)} \bigl(
e^{- r \tau h}\bigl(K-S_0 e^{\hat{X}^{(n)}_{\tau}} \bigr)^{+}
\bigr).
\]
In view of Theorem \ref{thmm2.1} and Remark \ref{rem2.2},
we directly conclude that
\[
\lim_{n\rightarrow\infty}V^{(n)}=V.
\]

Next, we describe a dynamical programming
algorithm for the calculation of~$V^{(n)}$.
Observe that for a
given $k \in\{0,\ldots,n\}$
the random variables $X^{(n)}_k$ and $Y^{(n)}_k$
take values on the grid
\begin{eqnarray*}
&& x_0+(2l-k)\sqrt{\eta h},\qquad 0 \leq l \leq k,
\\
&& y_0+(2m-k)\sqrt{\eta\bigl(1-\rho^2\bigr)h },\qquad  0 \leq m
\leq k,
\end{eqnarray*}
respectively.
For nonnegative integers $ m, l \le k \le n$ and
$\xi_x, \xi_y \in\{-1,+1\}$, let
\[
V^{(n)}_k(l,m,\xi_x,\xi_y)
\]
be the value of the option
at time $k$ when the Markov process
is given by
\begin{eqnarray*}
&&Z\Xi_k= \bigl(X^{(n)}_k, Y^{(n)}_k,
\xi^X_k, \xi^Y_k\bigr) =
F_k(l,m,\xi_x,\xi_y)
\\
&&\hspace*{20pt}:= \bigl(x_0+(2l-k)\sqrt{\eta h}, y_0+(2m-k)\sqrt {
\eta\bigl(1-\rho^2\bigr)h }, \xi_x, \xi_y
\bigr).
\end{eqnarray*}
The above function $F_k$ is invertible
with an inverse $F_k^{-1}$.
We sometimes,
with an abuse of notation, write
\[
V^{(n)}_{k-1}(\Xi) = V^{(n)}\bigl(
F_{k-1}^{-1}(\Xi)\bigr)
\]
for any four tuple $\Xi$
given by $F_{k-1}(l,m,\xi_x,\xi_y)$
for some $(l,m,\xi_x,\xi_y)$.
With this convention,
it is not straightforward to state
the dynamic programming equation
(see, e.g., \cite{PS}, Chapter~1),
%
\begin{equation}
\label{e.putdpp}\quad V^{(n)}_{k-1}(\Xi) = \max \bigl\{ \bigl( K -
S_0\exp(\hat X_{k-1}) \bigr)^+, \mathbb{E}^{(n)}
\bigl[ V^{(n)}_{k}(\Xi_{k}) {|} \Xi_{k-1}
=\Xi \bigr] \bigr\}.
\end{equation}

We continue by rewriting
the dynamic programming
equation in an algorithmic
manner.
In view of (\ref{alphabetadefinition})--(\ref{ksiYprob}),
for any $1\leq k\leq n$ and
$0\leq l,m \leq k-1$, we define
\begin{eqnarray*}
\mathcal{X}_{k} &:=& x_0+( 2l-k )\sqrt{ \eta h},
\\
\mathcal{Y}_{k} &:=& y_0+( 2m-k )\sqrt{ \eta\bigl(1-
\rho^2\bigr) h},
\end{eqnarray*}
where both of the above are functions of
$(l,m)$, but this dependence is suppressed in the notation.
Similarly, we define two probabilities
\begin{eqnarray*}
p_k(l,m,\xi_x,\xi_y)&:=&
\frac{\exp( rh + \sqrt{\eta h} \Psi
_{k-1} \xi
_x ) - \exp(-\sqrt{\eta h} \Psi_{k} )} {
\exp(\sqrt{\eta h} \Psi_{k} ) - \exp(-\sqrt{\eta h} \Psi_{k}
)},
\\
 q_k(l,m,\xi_x,\xi_y)&:=& \biggl( \min
\biggl\{1, \frac{1}2 + \frac{\alpha_{k-1}( l-\xi_x,m-\xi
_y) \xi
_y } {
2 \Psi_{k} } + \frac{\sqrt{h} \mu_{y,k} }{2 \sqrt{\eta(1-\rho^2)} \Psi_{k}} \biggr\}
\biggr)^{+},
\end{eqnarray*}
where $\alpha^{(n)}_0\equiv0$ and
\begin{eqnarray*}
\alpha_k(l,m) &:= & \frac{\max ( A_n, \sigma^2
( \mathcal{X}_{k-1}, \mathcal{Y}_{k-1}  )  )
-1}{2},
\\
\Psi_{k} &:=& 1+\alpha^{(n)}_{k}(l,m),
\\
\mu_{y,k} &:=& \mu_y (\mathcal{X}_{k-1},
\mathcal{Y}_{k-1}).
\end{eqnarray*}
As we remarked earlier, in our actual
numerical codes, we simply define
$\alpha= (\sigma^2 -1)/2$ without the
truncation with $A_n$ and instead truncate
$p_k$, above, to ensure that it stays within the unit
interval.

Observe that
\begin{eqnarray*}
p_k(l,m,\xi_x,\xi_y)&=&
\mathbb{P}^{(n)} \bigl( \xi^X_k=1 |
\Xi_{k-1}= F_{k-1}(l,m,\xi_x,\xi_y)
\bigr),
\\
q_k(l,m,\xi_x,\xi_y)&=&
\mathbb{P}^{(n)} \bigl( \xi^Y_k=1 |
\Xi_{k-1}= F_{k-1}(l,m,\xi_x,\xi_y)
\bigr).
\end{eqnarray*}
Moreover,
\[
\mathbb{P}^{(n)}_{k-1} \bigl( \xi^X_k=1,
\xi^Y_k=1 \bigr) = p_k(l,m,
\xi_x,\xi_y) q_k(l,m,\xi_x,
\xi_y).
\]
One can easily obtain expressions for the
other three probabilities as well.

We are now ready to restate the
dynamic programming equation \reff{e.putdpp}.
Indeed, $V^{(n)}_k(l,m,\xi_x,\xi_y)$
is the unique solution of the
following recursive relations:
%
%
\[
V^{(n)}_n(l,m,\xi_x,\xi_y)= \bigl( K- \exp{
(\mathcal{X}_{n}+ \sqrt{\eta h}\alpha_n \xi_x )} \bigr)^{+},
\]
and for $1\leq k \leq n$,
\[
V^{(n)}_{k-1}(l,m,\xi_x,\xi_y)=\max\bigl\{
\bigl( K-\exp(\mathcal{X}_{k-1}+\sqrt{\eta h}\alpha_{k-1} \xi_x)\bigr)^+,
\mathcal{E} \bigl(V^{(n)}_{k}\bigr)\bigr\},
\]
%
where
\begin{eqnarray*}
&&\Ec\bigl(V^{(n)}_{k}\bigr) \\
&&\qquad= \mathbb{E}^{(n)}
\bigl[ V^{(n)}_{k}(\Xi_{k}) | \Xi_{k-1}
=F_{k-1}(l,m,\xi_x,\xi_y) \bigr]
\\
&&\qquad=\sum_{i,j=0}^{1}
\mathbb{P}^{(n)}_{k-1} \bigl( \xi^X_k=2i-1,
\xi^Y_k=2j-1 \bigr) V^{(n)}_k(l+i,m+j,2i-1,2j-1)
\\
&&\qquad= \sum_{i,j=0}^{1}
\bigl[1-i+(2i-1) p_k(l,m,\xi_x,\xi_y)\bigr]
\bigl[1-j+(2j-1)q_k(l,m,\xi_x,\xi_y)\bigr]
\\
&&\hspace*{20pt}\qquad\quad{}\times V^{(n)}_k(l+i,m+j,2i-1,2j-1).
\end{eqnarray*}

Then our approximation is simply given by
\[
V_n=V^{(n)}_0(0,0,0,0).
\]

\subsection{Lookback options}
Consider a lookback put option with a fixed strike $K$,
that is, an option with payoff
$(K-\min_{0\leq t\leq T}S_t)^{+}$.
Again, we want to approximate the price
\[
\hat V=\mathbb{E} \Bigl(e^{-rT} \Bigl(K-\min_{0\leq t\leq T}S_t
\Bigr)^{+} \Bigr).
\]
Since the running minimum of the processes
\[
{\bigl\{\exp\bigl(X^{(n)}_k\bigr)\bigr\}}_{k=0}^n,\qquad
n\in\mathbb{N}
\]
lies on a grid, we will use these processes instead of the martingale
$\exp(\hat X^{(n)}_k)$.
The advantage of the processes $\exp(X^{(n)}_k)$
becomes clear
when we describe the
dynamical programming algorithm below.

We set
%
\begin{equation}
\label{3.3} \hat V^{(n)}=\mathbb{E}^{(n)}
\Bigl(e^{-rT} \Bigl(K-S_0\exp \Bigl(\min
_{0\leq i\leq n} {X}^{(n)}_{i} \Bigr)
\Bigr)^{+} \Bigr).
\end{equation}
By Theorem \ref{thmm2.1} we conclude that
$\hat V^{(n)}$ converges to $\hat V$.

First, we observe that the random variable
\[
z_k:= \min_{0\leq i\leq k} \sum
_{j=1}^i \xi^X_j
\]
takes values on the grid $\{-k,1-k,\ldots,0\}$.

Using the notation and the conventions
of the previous subsection, for $0\leq k\leq n$, we let
$\hat V^{(n)}_k(l,m,z,\xi_x,\xi_y)$
to be the option price at time $k$.
The extra state variable $z$ denotes
the value of the running minimum $z_k$ at time $k$.
Then, $\hat V^{(n)}$ is the unique solution of
\[
\hat V^{(n)}_n(l,m,z,\xi_x,
\xi_y)= \bigl(K-S_0\exp (-\sqrt{\eta h} z )
\bigr)^{+},
\]
and for $1\leq k \leq n$,
\[
\hat V^{(n)}_{k-1}(l,m,z,\xi_x,
\xi_y)= \max \bigl\{ \bigl(K-S_0\exp (-\sqrt{\eta h} z )
\bigr)^+, \hat\Ec\bigl( V^{(n)}_{k}\bigr) \bigr\},
\]
where
\begin{eqnarray*}
\hat\Ec\bigl( V^{(n)}_{k}\bigr) &=& \sum
_{i,j=0}^{1} \mathbb{P}^{(n)}_{k-1}
\bigl( \xi^X_k=2i-1, \xi^Y_k=2j-1
\bigr)
\\
&&\hspace*{20pt}{} \times\hat V^{(n)}_k (l+i,m+j,z+\chi_{\{i=0,
z+2l=k-1\}},2i-1,2j-1
),
\end{eqnarray*}
and $\chi_{Q}$ is the characteristic set of $Q$.
Finally,
\[
\hat V_n=\hat V^{(n)}_0(0,0,0,0,0).
\]


\section{Proof of Theorem \texorpdfstring{\protect\ref{thmm2.1}}{4.1}}
\label{sec:5}
\setcounter{equation}{0}
In this section we provide a proof of Theorem \ref{thmm2.1}.
Our main tool is the martingale convergence result of
Theorem 7.4.1 in~\cite{EK}.

In view of
(\ref{Xtree})--(\ref{Yhattree}) and (\ref{alphabetadefinition}),
we have the following inequality
for all sufficiently
large $n$:
\begin{eqnarray*}
\bigl|\hat{X}^{(n)}_k\bigr|&\geq&\bigl|X^{(n)}_k\bigr|-
\tfrac{1}{3}\bigl(\bigl|X^{(n)}_{k}\bigr|+\bigl|Y^{(n)}_k\bigr|+1
\bigr),
\\
\bigl|\hat{Y}^{(n)}_k\bigr|&\geq&\bigl|Y^{(n)}_k\bigr|-
\tfrac{1}{3}\bigl(\bigl|X^{(n)}_{k}\bigr|+\bigl|Y^{(n)}_k\bigr|+1
\bigr).
\end{eqnarray*}
Therefore,
%
\begin{equation}
\label{5.0} \bigl|X^{(n)}_k\bigr|+\bigl|Y^{(n)}_k\bigr|
\leq3\bigl(\bigl|\hat X^{(n)}_k\bigr| +\bigl|\hat Y^{(n)}_k\bigr|+1
\bigr), \qquad k=0,1,\ldots,n.
\end{equation}

This together with (\ref{Xhattree})--(\ref{Yhattree}) and (\ref
{alphabetadefinition})
imply that there exists a constant $c>0$ satisfying
%
\begin{eqnarray}
\label{5.1} \bigl|X^{(n)}_k-\hat{X}^{(n)}_k\bigr|+\bigl|Y^{(n)}_k-
\hat Y^{(n)}_k\bigr| \leq\frac{ c(1+|\hat{X}^{(n)}_k|
+|\hat{Y}^{(n)}_k|)}{\sqrt{n}},
\nonumber
\\[-8pt]
\\[-8pt]
 \eqntext{k=0,1,\ldots,n.}
\end{eqnarray}
It is sufficient to establish that
%
\begin{equation}
\label{new} \bigl\{\bigl(\hat X^{(n)}_{[nt/T]},\hat
Y^{(n)}_{[nt/T]}\bigr)\bigr\}_{t=0}^T
\Rightarrow{\bigl\{(x_t,y_t)\bigr\}}_{t=0}^T.
\end{equation}
Indeed, from (\ref{5.1}) it follows that
\begin{eqnarray*}
\hat{X}^{(n)}_k- \frac{ c(1+|\hat{X}^{(n)}_k|
+|\hat{Y}^{(n)}_k|)}{\sqrt{n}}&\leq&
X^{(n)}_k\leq\hat{X}^{(n)}_k+
\frac
{ c(1+|\hat{X}^{(n)}_k|
+|\hat{Y}^{(n)}_k|)}{\sqrt{n}},
\\
\hat{Y}^{(n)}_k- \frac{ c(1+|\hat{X}^{(n)}_k|
+|\hat{Y}^{(n)}_k|)}{\sqrt{n}}&\leq&
Y^{(n)}_k\leq\hat{Y}^{(n)}_k+
\frac
{ c(1+|\hat{X}^{(n)}_k|
+|\hat{Y}^{(n)}_k|)}{\sqrt{n}}.
\end{eqnarray*}
From (\ref{new}) it follows that
the sequences
\begin{eqnarray*}
\hspace*{-6pt}&&\biggl\{\hspace*{-1.3pt} \biggl(\hat{X}^{(n)}_{[nt/T]}\hspace*{-0.5pt}-\hspace*{-0.5pt} \frac{ c(1+|\hat{X}^{(n)}_{[nt/T]}|
+|\hat{Y}^{(n)}_{[nt/T]}|)}{\sqrt{n}},
\hat{Y}^{(n)}_{[nt/T]}\hspace*{-0.5pt}-\hspace*{-0.5pt} \frac{ c(1+|\hat{X}^{(n)}_{[nt/T]}|
+|\hat{Y}^{(n)}_{[nt/T]}|)}{\sqrt{n}} \biggr)\hspace*{-1.3pt} \biggr\}\hspace*{-0.5pt},
\\
\hspace*{-6pt}&&\biggl\{ \hspace*{-1.3pt}\biggl(\hat{X}^{(n)}_{[nt/T]}\hspace*{-0.5pt}+\hspace*{-0.5pt} \frac{ c(1+|\hat{X}^{(n)}_{[nt/T]}|
+|\hat{Y}^{(n)}_{[nt/T]}|)}{\sqrt{n}},
\hat{Y}^{(n)}_{[nt/T]}\hspace*{-0.5pt}+\hspace*{-0.5pt}\frac{ c(1+|\hat{X}^{(n)}_{[nt/T]}|
+|\hat{Y}^{(n)}_{[nt/T]}|)}{\sqrt{n}} \biggr)\hspace*{-1.3pt} \biggr\}
\end{eqnarray*}
converge weakly to ${\{(x_t,y_t)\}}_{t=0}^T$.
Thus Theorem \ref{thmm2.1}
follows from (\ref{new}).
For any $0\leq k \leq n$, set
\begin{eqnarray*}
\label{5.2} A^{n,x}_k &= &\sum
_{j=1}^k \mathbb{E}^{(n)}_{j-1}
\bigl(\hat{X}^{(n)}_j-\hat{X}^{(n)}_{j-1}
\bigr), \qquad A^{n,y}_k = \sum_{j=1}^k
\mathbb{E}^{(n)}_{j-1} \bigl(\hat{Y}^{(n)}_j-
\hat{Y}^{(n)}_{j-1} \bigr),
\\
M^{n,x}_k &=& \hat{X}^{(n)}_k-A^{n,x}_k,\qquad
M^{n,y}_k = \hat{Y}^{(n)}_k-A^{n,y}_k,
\\
A^{n,x,x}_k &=& \sum_{j=1}^k
\mathbb{E}^{(n)}_{j-1} \bigl( \bigl(M^{n,x}_j-M^{n,x}_{j-1}
\bigr)^2 \bigr), \qquad A^{n,y,y}_k = \sum
_{j=1}^k\mathbb{E}^{(n)}_{j-1}
\bigl( \bigl(M^{n,y}_j-M^{n,y}_{j-1}
\bigr)^2 \bigr),
\\
A^{n,x,y}_k &=& \sum_{j=1}^k
\mathbb{E}^{(n)}_{j-1} \bigl( \bigl(M^{n,x}_j-M^{n,x}_{j-1}
\bigr) \bigl(M^{n,y}_j-M^{n,y}_{j-1}
\bigr) \bigr).
\end{eqnarray*}
Notice that the processes
$A^{n,x},A^{n,y},A^{n,x,x},A^{n,y,y},A^{n,x,y}$
are predictable, and the processes
$M^{n,x}, M^{n,y}$ are martingales.

We now fix a large $N>0$
and
define the stopping times by
\[
\sigma_n=\min\bigl\{k\dvtx \bigl|\hat{X}^{(n)}_k\bigr|+ \bigl|
\hat{Y}^{(n)}_k\bigr|\geq N\bigr\}\wedge n,\qquad n\in\mathbb{N}.
\]

Using \reff{Xtree}, \reff{Ytree} and \reff{5.1},
we conclude that
for all $k\leq\sigma_n$,
\[
\hat{X}^{(n)}_k-\hat{X}^{(n)}_{k-1}= O(1/
\sqrt n) \quad{\mbox{and}}\quad \hat{Y}^{(n)}_k-\hat{Y}^{(n)}_{k-1}=
O(1/\sqrt n),
\]
where in this section
$o(\cdot)$ and $O(\cdot)$
are defined uniformly in space, that is,
$O(1/\sqrt n)$ is a function
which is bounded by a deterministic constant
over $\sqrt n$, and $\sqrt{n} o(1/\sqrt n)$
converges uniformly to zero as $n$ tends to infinity.

By Theorem
7.4.1 in \cite{EK},
\reff{new} would result from
the following relations:
%
\begin{eqnarray}
\label{5.3} \lim_{n\rightarrow\infty} \max_{1\leq k\leq\sigma_n}
\Biggl\llvert A^{n,x}_k- h \sum_{i=0}^{k-1}
\mu_x\bigl(\hat X^{(n)}_{i},\hat
Y^{(n)}_{i}\bigr) \Biggr\rrvert &=& 0 \qquad\mbox{a.s.},
\\
\label{5.4} \lim_{n\rightarrow\infty} \max_{1\leq k\leq\sigma_n}
\Biggl\llvert A^{n,y}_k- h \sum_{i=0}^{k-1}
\mu_y\bigl(\hat X^{(n)}_{i},\hat
Y^{(n)}_{i}\bigr) \Biggr\rrvert& =& 0\qquad \mbox{a.s.},
\\
\label{5.5} \lim_{n\rightarrow\infty} \max_{1\leq k\leq\sigma_n}
\Biggl\llvert A^{n,x,x}_k- \eta h \sum
_{i=0}^{k-1} \sigma^2\bigl(\hat
X^{(n)}_{i},\hat Y^{(n)}_{i}\bigr) \Biggr
\rrvert &=& 0\qquad \mbox{a.s.},
\\
\label{5.6} \lim_{n\rightarrow\infty} \max_{1\leq k\leq\sigma_n}
\Biggl\llvert A^{n,y,y}_k-\eta\bigl(1-\rho^2\bigr)
h \sum_{i=0}^{k-1} \sigma^2\bigl(
\hat X^{(n)}_{i},\hat Y^{(n)}_{i}\bigr)
\Biggr\rrvert &=& 0 \qquad\mbox{a.s.},
\\
\label{5.7} \lim_{n\rightarrow\infty} \max_{1\leq k\leq\sigma_n}
\bigl|A^{n,x,y}_k\bigr| &=& 0\qquad \mbox{a.s.}
\end{eqnarray}

The rest of the proof is devoted to
the verification of the above identities.

We start with a proof of \reff{5.3}.
Since $\sigma^2(x,y)$ is Lipschitz continuous,
\reff{Xtree}, \reff{Ytree} and \reff{alphabetadefinition}
imply that
\[
|\hat\alpha_k-\hat\alpha_{k-1}|=O(\sqrt h).
\]
In view of \reff{5.0},
for $k<\sigma_n$, we have
\[
-\tfrac{1}{2}\leq\alpha_k\leq\hat c (N+1)
\]
for some constant $\hat c$.
Since the event $ k < \sigma_n$ is
$\mathcal{F}_{k-1}$-measurable,
\[
\mathbb{P}^{(n)} \bigl(\xi^X_{k}=1 {\mbox{ and }}
k< \sigma_n | \Xi _{k-1} \bigr) = \chi_{ \{ k < \sigma_n \} }
\mathbb{P}^{(n)} \bigl(\xi^X_{k}=1 | \Xi
_{k-1} \bigr) = \chi_{ \{ k < \sigma_n \} } p_k.
\]
We now use
the above estimates,
the definition \reff{ksiXprob} of
the transition probability
$p_k$
and Taylor expansion. Then, on the set $k<\sigma_n$,
%
\begin{eqnarray}
\label{e.pxest} &&\mathbb{P}^{(n)} \bigl(\xi^X_{k}=1
| \Xi_{k-1} \bigr)
\nonumber
\\
&&\qquad = \frac{r h + \sqrt{\eta h } (1+\hat\alpha_{k-1}\xi^X_{k-1}+\hat
\alpha
_k)-\eta h (1/2+\hat\alpha_k)+o(h)} {
2 \sqrt{\eta h } (1+\hat\alpha_k)+o(h)}
\nonumber
\\[-8pt]
\\[-8pt]
\nonumber
&&\qquad = \frac{ r h + \sqrt{\eta h } (1+\hat\alpha_{k-1}\xi^X_{k-1}
+\hat\alpha_k)- \eta h (1/2+\hat\alpha_k)} {
2 \sqrt{\eta h } (1+\hat\alpha_k)}+ o(\sqrt h)
\\
\nonumber&&\qquad =\frac{1}2 + \frac{\hat\alpha_{k-1}}{2(1+\hat\alpha_{k})} \xi^X_{k-1} +
\frac{rh - \eta(1/2+ \hat\alpha_k) h}{2(1+\hat\alpha_{k})} + o(h).
\end{eqnarray}\eject\noindent
We thus conclude that on the event $k< \sigma_n$,
the following estimate holds:
\begin{eqnarray*}
&&\mathbb{E}^{(n)}_{k-1} \bigl[\hat{X}^{(n)}_k-
\hat {X}^{(n)}_{k-1} \bigr] \\
&&\qquad= \sqrt{\eta h }
\mathbb{E}^{(n)}_{k-1} \bigl[ (1+\hat\alpha_k)
\xi^X_k - \hat\alpha_{k-1} \xi^X_{k-1}
\bigr]
\\
&&\qquad= \sqrt{\eta h } \bigl[(1+\hat\alpha_k) \bigl(2
\mathbb{P}^{(n)} \bigl(\xi^X_{k}=1 |
\Xi_{k-1}\bigr)-1\bigr) -\hat\alpha_{k-1}\xi^X_{k-1}
\bigr]
\\
&&\qquad= r h -\eta \bigl(\tfrac{1}{2}+\hat\alpha_k \bigr)h+o(h)
\\
&&\qquad= \mu_x\bigl(\hat X^{(n)}_{k-1},\hat
Y^{(n)}_{k-1}\bigr)h + o(h),
\end{eqnarray*}
where the last equality follows from
the definition of $\hat\alpha$,
the Lipschitz continuity of $\mu(x,y)$ and
\reff{5.1}. Then
\reff{5.3} follows directly
from the above estimate.

We continue with a proof of \reff{5.4}.
We start with the definition
of $q_k$
and use the truncation introduced in
\reff{alphabetadefinition}. On
$k<\sigma_n$, this fields
the following estimate:
\[
2 \times\mathbb{P}^{(n)} \bigl(\xi^Y_{k}=1 |
\Xi_{k-1} \bigr)-1 = \frac{\hat\alpha_{k-1}}{1+\hat\alpha_{k}}\xi^Y_{k-1}+
\frac{\sqrt{h} \mu_y(X^{(n)}_{k-1},Y^{(n)}_{k-1})}{\sqrt{\eta
(1-\rho
^2)} (1+\hat\alpha_k)}.
\]
As before we directly estimate the on $k-1\leq\sigma_n$,
\begin{eqnarray*}
&&\mathbb{E}^{(n)}_{k-1} \bigl(\hat{Y}^{(n)}_k-
\hat {Y}^{(n)}_{k-1} \bigr)
\\
&&\qquad= \sqrt{\eta\bigl(1-\rho^2\bigr) h} \bigl((1+\hat
\alpha_k) \bigl(2\mathbb {P}^{(n)} \bigl(
\xi^Y_{k}=1 | \Xi_{k-1} \bigr) -1 \bigr)-\hat
\alpha_k\xi^Y_{k-1} \bigr)
\\
&&\qquad = \mu_y\bigl(\hat X^{(n)}_{k-1},\hat
Y^{(n)}_{k-1}\bigr)h+o(h).
\end{eqnarray*}
Again, the last equality follows from
\reff{5.1} and the fact that $\mu_y(x,y)$ is Lipschitz continuous.
This completes the proof of \reff{5.4}.

We continue with the quadratic estimates. Indeed,
by \reff{e.pxest}, on $k<\sigma_n$,
\[
2 \times\mathbb{P}^{(n)} \bigl(\xi^X_{k}=1|
\Xi_{k-1} \bigr)-1= \frac{\hat\alpha_{k-1}}{1+\hat\alpha_{k}}\xi^X_{k-1}+o(
\sqrt h).
\]
Since $A^{n,x}$ is predictable, on $k< \sigma_n$,
\[
\mathbb{E}^{(n)}_{k-1} \bigl( \bigl(M^{n,x}_k-M^{n,x}_{k-1}
\bigr)^2 \bigr) =\mathbb{E}^{(n)}_{k-1} \bigl(
\bigl(\hat{X}^{(n)}_k-\hat {X}^{(n)}_{k-1}
\bigr)^2 \bigr)+o(h)
\]
and
\begin{eqnarray*}
\hspace*{-4pt}&&\mathbb{E}^{(n)}_{k-1} \bigl( \bigl(\hat{X}^{(n)}_k-
\hat {X}^{(n)}_{k-1} \bigr)^2 \bigr)
\\
\hspace*{-8pt}&&\!\!\!\qquad=\eta h \bigl((1+\hat\alpha_k)^2+(\hat
\alpha_{k-1})^2 -2\hat\alpha_{k-1}(1+\hat
\alpha_k)\xi^X_{k-1} \bigl(2\mathbb{P}^{(n)}
\bigl(\xi^X_{k}=1 | \Xi_{k-1} \bigr)-1\bigr)
\bigr)
\\
\hspace*{-8pt}&&\!\!\!\qquad = \eta h \bigl(1+2\hat\alpha^{(n)}_k\bigr)
\\
\hspace*{-8pt}&&\!\!\!\qquad = \eta h \sigma^2\bigl(\hat X^{(n)}_{k-1},\hat
Y^{(n)}_{k-1}\bigr),
\end{eqnarray*}
and \reff{5.5} follows. Relation \reff{5.6} is proved
similarly.\vadjust{\goodbreak}

It remains to establish \reff{5.7}.
The processes $A^{n,x},A^{n,y}$ are predictable.
Thus, from \reff{independence}
it follows that, on $k< \sigma_n$,
\begin{eqnarray*}
&&\mathbb{E}^{(n)}_{k-1} \bigl( \bigl(M^{n,x}_k-M^{n,x}_{k-1}
\bigr) \bigl(M^{n,y}_k-M^{n,y}_{k-1}
\bigr) \bigr)
\\
&&\qquad= \mathbb{E}^{(n)}_{k-1} \bigl( \bigl(\hat{X}^{(n)}_{k}-
\hat{X}^{(n)}_{k-1} \bigr) \bigl(\hat{Y}^{(n)}_{k}-
\hat{Y}^{(n)}_{k-1} \bigr) \bigr)+o(h)
\\
&&\qquad=\mathbb{E}^{(n)}_{k-1} \bigl(\hat{X}^{(n)}_{k}-
\hat{X}^{(n)}_{k-1} \bigr) \mathbb{E}^{(n)}_{k-1}
\bigl(\hat{Y}^{(n)}_{k}-\hat {Y}^{(n)}_{k-1}
\bigr)+o(h)
\\
&&\qquad = o(h),
\end{eqnarray*}
where we used the fact that $\xi^X_{k}$ and $\xi^Y_{k}$ are
conditionally independent.

\section{Numerical results}

In this section, we present numerical results from our model for
European and American vanilla, lookback, geometric
and arithmetic Asian options under the Heston dynamics.
Our computations are obtained by a direct
implementation of the methodology described
in the previous sections. In particular,
we explicitly refrained from using known
numerical techniques that improve
the performance of the trees. This is done
to ensure the replicability of our reported
results.

\subsection{Vanillas}

In Tables~\ref{HestonEput}, \ref{HestonEcall} and \ref{NBput}, we use
the same parameter sets as in
Beliaeva and Nawalkha \cite{bn}, that is,
for European call and put options: strike $K =100$; initial stock prices:
$S_0 = 90, 95, 100, 105, 110$; maturities: $T =$ 1 month, 3 months and
6 months;
initial volatility values: $\sqrt{\nu_0} = 0.2, 0.3, 0.4$;
interest rate: $r = 0.05$; vol of vol: $\eta= 0.1$;
mean reversion rate: $\kappa= 3$; long run vol: $\theta= 0.04 $;
and correlation: $\rho= -0.7$. For American put options:
$K=100$, $S_0 = 90, 100, 110$; $T =$ 1 month, 3 months and 6 months;
$\sqrt{\nu_0} = 0.2, 0.4$; $\rho= -0.1, -0.7$; $r=0.05$; $\eta= 0.1$;
$\kappa= 3$, $\theta=0.04$.

\begin{table}
\tabcolsep=0pt
\caption{Convergence of European put prices versus analytical solution
of Heston \cite{hes}.
Parameters: $K = 100$, $r = 0.05$, $\eta= 0.1$, $\kappa= 3.0$,
$\theta= 0.04$ and $\rho= -0.7$}
\label{HestonEput}
{\fontsize{8.5}{10.5}{\selectfont
\begin{tabular*}{\textwidth}{@{\extracolsep{\fill}}lcd{1.4}d{2.4}d{2.4}d{2.4}d{2.4}d{2.2}d{2.2}d{2.2}@{}}
\hline
&  & & \multicolumn{3}{c}{\textbf{Tree}} &
&\multicolumn{3}{c@{}}{\textbf{Error \%}} \\[-6pt]
&  & & \multicolumn{3}{c}{\hrulefill} &
&\multicolumn{3}{c@{}}{\hrulefill} \\
\multicolumn{1}{@{}l}{$\bolds{S(0)}$}&
\multicolumn{1}{c}{{$\bolds{\sqrt{\nu_0}}$}}&
\multicolumn{1}{c}{{$\bolds{T}$}}&
\multicolumn{1}{c}{$\bolds{N = 200}$} &
\multicolumn{1}{c}{$\bolds{N = 350}$} &
\multicolumn{1}{c}{$\bolds{N = 500}$} &
\multicolumn{1}{c}{\multirow{2}{40pt}[10pt]{\centering\textbf{Analytical solution}}} &
\multicolumn{1}{c}{$\bolds{N = 200}$} &
\multicolumn{1}{c}{$\bolds{N = 350}$} & \multicolumn{1}{c@{}}{$\bolds{N = 500}$} \\
\hline
\phantom{0}90 & 0.2 & 0.0833 & 9.6541 & 9.6533 & 9.6533 & 9.6533 & 0.01 & 0.00 &
0.00 \\
\phantom{0}95 & 0.2 & 0.0833 & 5.2059 & 5.2084 & 5.2077 & 5.2074 & -0.03 & 0.02 &
0.01 \\
100 & 0.2 & 0.0833 & 2.0953 & 2.0960 & 2.0965 & 2.0971 & -0.08 & -0.05
& -0.03 \\
105 & 0.2 & 0.0833 & 0.6082 & 0.6047 & 0.6050 & 0.6053 & 0.48 & -0.10 &
-0.06 \\
110 & 0.2 & 0.0833 & 0.1267 & 0.1271 & 0.1270 & 0.1265 & 0.11 & 0.48 &
0.35 \\
\phantom{0}90 & 0.3 & 0.0833 & 9.9913 & 9.9900 & 9.9900 & 9.9905 & 0.01 & 0.00 &
0.00 \\
\phantom{0}95 & 0.3 & 0.0833 & 6.0147 & 6.0170 & 6.0162 & 6.0155 & -0.01 & 0.02 &
0.01 \\
100 & 0.3 & 0.0833 & 3.1308 & 3.1288 & 3.1290 & 3.1302 & 0.02 & -0.05 &
-0.04 \\
105 & 0.3 & 0.0833 & 1.4001 & 1.3955 & 1.3955 & 1.3967 & 0.25 & -0.08 &
-0.09 \\
110 & 0.3 & 0.0833 & 0.5365 & 0.5374 & 0.5372 & 0.5367 & -0.05 & 0.13 &
0.09 \\
\phantom{0}90 & 0.4 & 0.0833 & 10.5687 & 10.5670 & 10.5668 & 10.5668 & 0.02 & 0.00
& 0.00 \\
\phantom{0}95 & 0.4 & 0.0833 & 6.9357 & 6.9363 & 6.9352 & 6.9335 & 0.03 & 0.04 &
0.02 \\
100 & 0.4 & 0.0833 & 4.1893 & 4.1864 & 4.1861 & 4.1852 & 0.10 & 0.03 &
0.02 \\
105 & 0.4 & 0.0833 & 2.3280 & 2.3232 & 2.3229 & 2.3222 & 0.25 & 0.04 &
0.03 \\
110 & 0.4 & 0.0833 & 1.1893 & 1.1897 & 1.1893 & 1.1882 & 0.09 & 0.13 &
0.09 \\
\phantom{0}90 & 0.2 & 0.25 & 9.5736 & 9.5693 & 9.5694 & 9.5698 & 0.04 & 0.00 &
0.00 \\
\phantom{0}95 & 0.2 & 0.25 & 5.9691 & 5.9685 & 5.9693 & 5.9692 & 0.00 & -0.01 &
0.00 \\
100 & 0.2 & 0.25 & 3.3742 & 3.3774 & 3.3794 & 3.3770 & -0.08 & 0.01 &
0.07 \\
105 & 0.2 & 0.25 & 1.7420 & 1.7393 & 1.7402 & 1.7410 & 0.06 & -0.10 &
-0.05 \\
110 & 0.2 & 0.25 & 0.8290 & 0.8249 & 0.8253 & 0.8259 & 0.37 & -0.13 &
-0.08 \\
\phantom{0}90 & 0.3 & 0.25 & 10.5941 & 10.5879 & 10.5882 & 10.5893 & 0.04 & -0.01
& -0.01 \\
\phantom{0}95 & 0.3 & 0.25 & 7.3343 & 7.3327 & 7.3329 & 7.3316 & 0.04 & 0.02 &
0.02 \\
100 & 0.3 & 0.25 & 4.8279 & 4.8331 & 4.8340 & 4.8310 & -0.06 & 0.04 &
0.06 \\
105 & 0.3 & 0.25 & 3.0420 & 3.0379 & 3.0391 & 3.0388 & 0.11 & -0.03 &
0.01 \\
110 & 0.3 & 0.25 & 1.8368 & 1.8320 & 1.8319 & 1.8325 & 0.23 & -0.03 &
-0.03 \\
\phantom{0}90 & 0.4 & 0.25 & 11.8375 & 11.8281 & 11.8288 & 11.8287 & 0.07 & 0.00 &
0.00 \\
\phantom{0}95 & 0.4 & 0.25 & 8.8120 & 8.8081 & 8.8070 & 8.8035 & 0.10 & 0.05 &
0.04 \\
100 & 0.4 & 0.25 & 6.3762 & 6.3790 & 6.3786 & 6.3735 & 0.04 & 0.09 &
0.08 \\
105 & 0.4 & 0.25 & 4.5066 & 4.5005 & 4.5004 & 4.4976 & 0.20 & 0.06 &
0.06 \\
110 & 0.4 & 0.25 & 3.1099 & 3.1035 & 3.1025 & 3.1011 & 0.28 & 0.08 &
0.05 \\
\phantom{0}90 & 0.2 & 0.5 & 9.7547 & 9.7545 & 9.7606 & 9.7572 & -0.03 & -0.03 &
0.04 \\
\phantom{0}95 & 0.2 & 0.5 & 6.7258 & 6.7248 & 6.7185 & 6.7199 & 0.09 & 0.07 &
-0.02 \\
100 & 0.2 & 0.5 & 4.4355 & 4.4369 & 4.4320 & 4.4312 & 0.10 & 0.13 &
0.02 \\
105 & 0.2 & 0.5 & 2.8077 & 2.8159 & 2.8100 & 2.8107 & -0.11 & 0.18 &
-0.02 \\
110 & 0.2 & 0.5 & 1.7286 & 1.7289 & 1.7275 & 1.7240 & 0.27 & 0.28 &
0.20 \\
\phantom{0}90 & 0.3 & 0.5 & 11.0786 & 11.0792 & 11.0845 & 11.0807 & -0.02 & -0.01
& 0.03 \\
\phantom{0}95 & 0.3 & 0.5 & 8.2445 & 8.2422 & 8.2367 & 8.2363 & 0.10 & 0.07 & 0.00
\\
100 & 0.3 & 0.5 & 5.9835 & 5.9830 & 5.9784 & 5.9763 & 0.12 & 0.11 &
0.04 \\
105 & 0.3 & 0.5 & 4.2450 & 4.2504 & 4.2449 & 4.2443 & 0.02 & 0.15 &
0.02 \\
110 & 0.3 & 0.5 & 2.9647 & 2.9640 & 2.9623 & 2.9582 & 0.22 & 0.20 &
0.14 \\
\phantom{0}90 & 0.4 & 0.5 & 12.6195 & 12.6199 & 12.6231 & 12.6171 & 0.02 & 0.02 &
0.05 \\
\phantom{0}95 & 0.4 & 0.5 & 9.9373 & 9.9318 & 9.9260 & 9.9223 & 0.15 & 0.10 & 0.04
\\
100 & 0.4 & 0.5 & 7.7110 & 7.7069 & 7.7017 & 7.6965 & 0.19 & 0.13 &
0.07 \\
105 & 0.4 & 0.5 & 5.9065 & 5.9075 & 5.9015 & 5.8978 & 0.15 & 0.17 &
0.06 \\
110 & 0.4 & 0.5 & 4.4841 & 4.4806 & 4.4779 & 4.4716 & 0.28 & 0.20 &
0.14 \\
\hline
\end{tabular*}}}
\end{table}
%

\begin{table}
\tabcolsep=0pt
\caption{Convergence of European call prices versus analytical solution
of Heston \cite{hes}.
Parameters: $K = 100$, $r = 0.05$, $\eta= 0.1$, $\kappa= 3.0$,
$\theta= 0.04$, and $\rho= -0.7$}
\label{HestonEcall}
{\fontsize{8.5}{10.5}{\selectfont
\begin{tabular*}{\textwidth}{@{\extracolsep{\fill}}lcd{1.4}d{2.4}d{2.4}d{2.4}d{2.4}d{2.2}d{2.2}d{2.2}@{}}
\hline
&  & & \multicolumn{3}{c}{\textbf{Tree}} &
&\multicolumn{3}{c@{}}{\textbf{Error \%}} \\[-6pt]
&  & & \multicolumn{3}{c}{\hrulefill} &
&\multicolumn{3}{c@{}}{\hrulefill} \\
\multicolumn{1}{@{}l}{$\bolds{S(0)}$}&
\multicolumn{1}{c}{{$\bolds{\sqrt{\nu_0}}$}}&
\multicolumn{1}{c}{{$\bolds{T}$}}&
\multicolumn{1}{c}{$\bolds{N = 200}$} &
\multicolumn{1}{c}{$\bolds{N = 350}$} &
\multicolumn{1}{c}{$\bolds{N = 500}$} &
\multicolumn{1}{c}{\multirow{2}{40pt}[10pt]{\centering\textbf{Analytical solution}}} &
\multicolumn{1}{c}{$\bolds{N = 200}$} &
\multicolumn{1}{c}{$\bolds{N = 350}$} & \multicolumn{1}{c@{}}{$\bolds{N = 500}$} \\
\hline
\phantom{0}90 & 0.2 & 0.0833 & 0.0699 & 0.0691 & 0.0691 & 0.0691 & 1.13 & -0.05 &
0.02 \\
\phantom{0}95 & 0.2 & 0.0833 & 0.6217 & 0.6242 & 0.6235 & 0.6232 & -0.23 & 0.17 &
0.06 \\
100 & 0.2 & 0.0833 & 2.5111 & 2.5118 & 2.5122 & 2.5129 & -0.07 & -0.04
& -0.02 \\
105 & 0.2 & 0.0833 & 6.0240 & 6.0205 & 6.0208 & 6.0211 & 0.05 & -0.01 &
-0.01 \\
110 & 0.2 & 0.0833 & 10.5425 & 10.5429 & 10.5428 & 10.5423 & 0.00 &
0.01 & 0.00 \\
\phantom{0}90 & 0.3 & 0.0833 & 0.4071 & 0.4058 & 0.4058 & 0.4063 & 0.20 & -0.12 &
-0.12 \\
\phantom{0}95 & 0.3 & 0.0833 & 1.4305 & 1.4328 & 1.4320 & 1.4313 & -0.06 & 0.10 &
0.05 \\
100 & 0.3 & 0.0833 & 3.5466 & 3.5446 & 3.5448 & 3.5460 & 0.02 & -0.04 &
-0.04 \\
105 & 0.3 & 0.0833 & 6.8159 & 6.8113 & 6.8113 & 6.8125 & 0.05 & -0.02 &
-0.02 \\
110 & 0.3 & 0.0833 & 10.9523 & 10.9532 & 10.9530 & 10.9525 & 0.00 &
0.01 & 0.00 \\
\phantom{0}90 & 0.4 & 0.0833 & 0.9845 & 0.9828 & 0.9826 & 0.9826 & 0.19 & 0.02 &
0.00 \\
\phantom{0}95 & 0.4 & 0.0833 & 2.3515 & 2.3521 & 2.3510 & 2.3493 & 0.10 & 0.12 &
0.07 \\
100 & 0.4 & 0.0833 & 4.6051 & 4.6022 & 4.6019 & 4.6010 & 0.09 & 0.03 &
0.02 \\
105 & 0.4 & 0.0833 & 7.7438 & 7.7390 & 7.7387 & 7.7380 & 0.08 & 0.01 &
0.01 \\
110 & 0.4 & 0.0833 & 11.6051 & 11.6055 & 11.6051 & 11.6040 & 0.01 &
0.01 & 0.01 \\
\phantom{0}90 & 0.2 & 0.25 & 0.8158 & 0.8115 & 0.8116 & 0.8120 & 0.47 & -0.05 &
-0.05 \\
\phantom{0}95 & 0.2 & 0.25 & 2.2113 & 2.2107 & 2.2116 & 2.2114 & -0.01 & -0.03 &
0.01 \\
100 & 0.2 & 0.25 & 4.6164 & 4.6196 & 4.6216 & 4.6192 & -0.06 & 0.01 &
0.05 \\
105 & 0.2 & 0.25 & 7.9842 & 7.9815 & 7.9824 & 7.9832 & 0.01 & -0.02 &
-0.01 \\
110 & 0.2 & 0.25 & 12.0712 & 12.0671 & 12.0675 & 12.0682 & 0.03 & -0.01
& -0.01 \\
\phantom{0}90 & 0.3 & 0.25 & 1.8363 & 1.8301 & 1.8305 & 1.8316 & 0.26 & -0.08 &
-0.06 \\
\phantom{0}95 & 0.3 & 0.25 & 3.5766 & 3.5750 & 3.5751 & 3.5738 & 0.08 & 0.03 &
0.03 \\
100 & 0.3 & 0.25 & 6.0701 & 6.0753 & 6.0762 & 6.0732 & -0.05 & 0.03 &
0.05 \\
105 & 0.3 & 0.25 & 9.2842 & 9.2802 & 9.2813 & 9.2810 & 0.04 & -0.01 &
0.00 \\
110 & 0.3 & 0.25 & 13.0790 & 13.0742 & 13.0741 & 13.0747 & 0.03 & 0.00
& 0.00 \\
\phantom{0}90 & 0.4 & 0.25 & 3.0797 & 3.0703 & 3.0710 & 3.0709 & 0.29 & -0.02 &
0.00 \\
\phantom{0}95 & 0.4 & 0.25 & 5.0542 & 5.0503 & 5.0493 & 5.0457 & 0.17 & 0.09 &
0.07 \\
100 & 0.4 & 0.25 & 7.6184 & 7.6212 & 7.6208 & 7.6157 & 0.04 & 0.07 &
0.07 \\
105 & 0.4 & 0.25 & 10.7488 & 10.7428 & 10.7426 & 10.7399 & 0.08 & 0.03
& 0.03 \\
110 & 0.4 & 0.25 & 14.3521 & 14.3457 & 14.3447 & 14.3433 & 0.06 & 0.02
& 0.01 \\
\phantom{0}90 & 0.2 & 0.5 & 2.2237 & 2.2235 & 2.2296 & 2.2262 & -0.11 & -0.12 &
0.15 \\
\phantom{0}95 & 0.2 & 0.5 & 4.1948 & 4.1938 & 4.1875 & 4.1889 & 0.14 & 0.12 &
-0.03 \\
100 & 0.2 & 0.5 & 6.9045 & 6.9060 & 6.9010 & 6.9002 & 0.06 & 0.08 &
0.01 \\
105 & 0.2 & 0.5 & 10.2767 & 10.2849 & 10.2790 & 10.2797 & -0.03 & 0.05
& -0.01 \\
110 & 0.2 & 0.5 & 14.1976 & 14.1979 & 14.1965 & 14.1930 & 0.03 & 0.03 &
0.02 \\
\phantom{0}90 & 0.3 & 0.5 & 3.5476 & 3.5483 & 3.5535 & 3.5497 & -0.06 & -0.04 &
0.11 \\
\phantom{0}95 & 0.3 & 0.5 & 5.7135 & 5.7112 & 5.7057 & 5.7053 & 0.14 & 0.10 & 0.01
\\
100 & 0.3 & 0.5 & 8.4525 & 8.4520 & 8.4474 & 8.4453 & 0.09 & 0.08 &
0.03 \\
105 & 0.3 & 0.5 & 11.7140 & 11.7194 & 11.7140 & 11.7133 & 0.01 & 0.05 &
0.01 \\
110 & 0.3 & 0.5 & 15.4337 & 15.4330 & 15.4313 & 15.4272 & 0.04 & 0.04 &
0.03 \\
\phantom{0}90 & 0.4 & 0.5 & 5.0885 & 5.0889 & 5.0921 & 5.0861 & 0.05 & 0.06 & 0.12
\\
\phantom{0}95 & 0.4 & 0.5 & 7.4063 & 7.4008 & 7.3950 & 7.3913 & 0.20 & 0.13 & 0.05
\\
100 & 0.4 & 0.5 & 10.1800 & 10.1759 & 10.1707 & 10.1655 & 0.14 & 0.10 &
0.05 \\
105 & 0.4 & 0.5 & 13.3755 & 13.3765 & 13.3705 & 13.3668 & 0.07 & 0.07 &
0.03 \\
110 & 0.4 & 0.5 & 16.9532 & 16.9496 & 16.9469 & 16.9406 & 0.07 & 0.05 &
0.04 \\
\hline
\end{tabular*}}}
\end{table}
%

\begin{table}
\def\arraystretch{0.95}
\caption{Comparison of American put prices calculated with our method
and with the control variate technique of Beliaeva and Nawalkha \cite
{bn}. Parameters: $K = 100$, $r = 0.05$, $\eta=
0.1$, $\kappa= 3.0$, $\theta= 0.04$, and $\rho= -0.7$}
\label{NBput}
\begin{tabular*}{\textwidth}{@{\extracolsep{\fill}}ld{2.1}cd{1.4}d{2.4}d{2.4}d{2.2}@{}}
\hline
& &  & & \multicolumn{1}{c}{\textbf{Tree}} & \multicolumn{1}{c}{\textbf{Control variate}} & \\
\multicolumn{1}{@{}l}{{$\bolds{S(0)}$}} &\multicolumn{1}{c}{{$\bolds{\rho}$}} & \multicolumn{1}{c}{{$\bolds{\sqrt{ \nu_0}}$}}&
\multicolumn{1}{c}{{$\bolds{T}$}} & \multicolumn{1}{c}{$\bolds{N = 250}$} & \multicolumn{1}{c}{$\bolds{N=200}$} & \multicolumn{1}{c@{}}{\textbf{Difference \%}}\\
\hline
\phantom{0}90 & -0.1 & 0.2 & 0.0833 & 10.0000 & 10.0000 & 0.00 \\
100 & -0.1 & 0.2 & 0.0833 & 2.1236 & 2.1254 & -0.08 \\
110 & -0.1 & 0.2 & 0.0833 & 0.1090 & 0.1091 & -0.05 \\
\phantom{0}90 & -0.7 & 0.2 & 0.0833 & 10.0000 & 9.9997 & 0.00 \\
100 & -0.7 & 0.2 & 0.0833 & 2.1249 & 2.1267 & -0.08 \\
110 & -0.7 & 0.2 & 0.0833 & 0.1273 & 0.1274 & -0.07 \\
\phantom{0}90 & -0.1 & 0.4 & 0.0833 & 10.7123 & 10.7100 & 0.02 \\
100 & -0.1 & 0.4 & 0.0833 & 4.2194 & 4.2158 & 0.08 \\
110 & -0.1 & 0.4 & 0.0833 & 1.1666 & 1.1667 & -0.01 \\
\phantom{0}90 & -0.7 & 0.4 & 0.0833 & 10.6843 & 10.6804 & 0.04 \\
100 & -0.7 & 0.4 & 0.0833 & 4.2183 & 4.2140 & 0.10 \\
110 & -0.7 & 0.4 & 0.0833 & 1.1942 & 1.1939 & 0.02 \\
\phantom{0}90 & -0.1 & 0.2 & 0.25 & 10.1713 & 10.1706 & 0.01 \\
100 & -0.1 & 0.2 & 0.25 & 3.4729 & 3.4747 & -0.05 \\
110 & -0.1 & 0.2 & 0.25 & 0.7726 & 0.7736 & -0.13 \\
\phantom{0}90 & -0.7 & 0.2 & 0.25 & 10.1222 & 10.1206 & 0.02 \\
100 & -0.7 & 0.2 & 0.25 & 3.4790 & 3.4807 & -0.05 \\
110 & -0.7 & 0.2 & 0.25 & 0.8405 & 0.8416 & -0.13 \\
\phantom{0}90 & -0.1 & 0.4 & 0.25 & 12.1880 & 12.1819 & 0.05 \\
100 & -0.1 & 0.4 & 0.25 & 6.5023 & 6.4964 & 0.09 \\
110 & -0.1 & 0.4 & 0.25 & 3.0952 & 3.0914 & 0.12 \\
\phantom{0}90 & -0.7 & 0.4 & 0.25 & 12.1245 & 12.1122 & 0.10 \\
100 & -0.7 & 0.4 & 0.25 & 6.4989 & 6.4899 & 0.14 \\
110 & -0.7 & 0.4 & 0.25 & 3.1512 & 3.1456 & 0.18 \\
\phantom{0}90 & -0.1 & 0.2 & 0.5 & 10.6521 & 10.6478 & 0.04 \\
100 & -0.1 & 0.2 & 0.5 & 4.6531 & 4.6473 & 0.12 \\
110 & -0.1 & 0.2 & 0.5 & 1.6857 & 1.6832 & 0.15 \\
\phantom{0}90 & -0.7 & 0.2 & 0.5 & 10.5682 & 10.5637 & 0.04 \\
100 & -0.7 & 0.2 & 0.5 & 4.6691 & 4.6636 & 0.12 \\
110 & -0.7 & 0.2 & 0.5 & 1.7899 & 1.7874 & 0.14 \\
\phantom{0}90 & -0.1 & 0.4 & 0.5 & 13.3279 & 13.3142 & 0.10 \\
100 & -0.1 & 0.4 & 0.5 & 8.0231 & 8.0083 & 0.18 \\
110 & -0.1 & 0.4 & 0.5 & 4.5554 & 4.5454 & 0.22 \\
\phantom{0}90 & -0.7 & 0.4 & 0.5 & 13.2431 & 13.2172 & 0.20 \\
100 & -0.7 & 0.4 & 0.5 & 8.0204 & 7.9998 & 0.26 \\
110 & -0.7 & 0.4 & 0.5 & 4.6328 & 4.6201 & 0.27 \\
\hline
\end{tabular*}
\end{table}

Tables~\ref{HestonEput} and \ref{HestonEcall} show the convergence of European
put and call prices computed
by our method compared to the closed form solutions
of Heston \cite{hes}. In the European case,
one can calculate errors as
Heston's solution is available in closed
form. The option prices computed for the
number of time steps $N = 200$, $350$ and
$500$ illustrate very good convergence
to the closed form solutions as reported in Tables~\ref{HestonEput} and \ref{HestonEcall}.
Furthermore,
one can verify that the put-call parity holds exactly for option prices at
each of these time steps sizes.
Clearly, this is the outcome of the fact that
our price process in any step size is a martingale.

Table~\ref{NBput} reports the difference between the American put prices
obtained from our method and those obtained by the control
variate (CV) technique of \cite{bn}.
The table shows that our numbers are in good
agreement with those obtained
by the CV method. The first three largest differences between the
models are
$(0.27 \%, 0.26 \%, 0.22 \% )$, and on average
there is a difference of $0.10 \%$ per option.
We should point out to the reader that\vadjust{\goodbreak}
the CV technique
computes the value of the put option
via the formula
\[
\mbox{CV American Price} =\mbox{Tree American} + \mbox{(Closed Form Euro} - \mbox{Tree Euro}).
\]
According to Beliaeva and Nawalkha \cite{bn}, this method is
particularly useful for longer
maturity options.

Chockalingam and Muthuraman \cite{cm} develop a
partial differential equations (PDE)
based finite difference method to price
American options under stochastic volatility.
More specifically, they transform the
free boundary problem resulting from the
pricing of American options into a sequence
of fixed-boundary problems of European type.
The prices listed in Tables~\ref{PDEput1} and \ref{PDEput2} are
taken from \cite{cm} as a benchmark for our tree-based method.
The authors provide the values arising from the projected successive
over relaxation
(PSOR) method and the component-wise splitting (CS) method.
They state that other
PDE-based methods (see Ikonen and Toivanen \cite{IT}
for a detailed analysis) fall
between these two in terms of speed/accuracy and
ease of implementation. As test parameters, they use
the most common parameter values for American options
under the Heston dynamics in the PDE-based literature: $K = 10$, $r = 0.1$,
$\eta= 0.9$, $\kappa= 5.0$, $\theta= 0.16$ and $\rho= 0.1$, $T=0.25$,
$\sqrt{\nu_0} = 0.25, 0.5$. Following \cite{cm}, we take the
prices computed by Ikonen and Toivanen \cite{IT}
(using the CS method together
with a very fine grid) as the reference values.
From Tables~\ref{PDEput1} and \ref{PDEput2}, one can clearly
conclude that our results for both $N=250$ and $N=350$
are very close to reference values.

\begin{table}
\caption{American put prices determined with our tree approach and
finite difference methods. Parameters: $K = 10$, $r = 0.1$, $\eta=
0.9$, $\kappa= 5.0$, $\theta= 0.16$, and
$\rho= 0.1$, $T=0.25$, $\sqrt{\nu_0} = \textbf{0.25}$}
\label{PDEput1}
\begin{tabular*}{\textwidth}{@{\extracolsep{\fill}}lcccccc@{}}
\hline
& & \multicolumn{5}{c@{}}{$\bolds{S_0}$} \\[-6pt]
& & \multicolumn{5}{c@{}}{\hrulefill} \\
\textbf{Method} & \textbf{Grid size} & \textbf{8} & \textbf{9} & \textbf{10} & \textbf{11} & \textbf{12} \\
\hline
{PSOR} & (40, 16, 8) & 2.0000 & 1.0952 & 0.4966 & 0.2042 &
0.0838 \\
& (60, 32, 66) & 2.0000 & 1.1037 & 0.5142 & 0.2105 & 0.0815 \\
& (120, 64, 130) & 2.0000 & 1.1064 & 0.5182 & 0.2126 & 0.0819 \\
& (240, 128, 258) & 2.0000 & 1.1071 & 0.5193 & 0.2133 & 0.0820 \\[6pt]
Componentwise & (40, 16, 8) & 2.0004 & 1.1003 & 0.4991 & 0.2035 & 0.0828
\\
splitting & (60, 32, 66) & 2.0000 & 1.1043 & 0.5147 & 0.2104 & 0.0813 \\
& (120, 64, 130) & 2.0000 & 1.1066 & 0.5183 & 0.2126 & 0.0819 \\
& (240, 128, 258) & 2.0000 & 1.1073 & 0.5194 & 0.2133 & 0.0820 \\[6pt]
Transformation & (40, 16, 8) & 2.0000 & 1.0952 & 0.4966 & 0.2042 & 0.0838
\\
procedure & (60, 32, 66) & 2.0000 & 1.1035 & 0.5142 & 0.2105 & 0.0815 \\
& (120, 64, 130) & 2.0000 & 1.1063 & 0.5181 & 0.2126 & 0.0819 \\
& (240, 128, 258) & 2.0000 & 1.1071 & 0.5193 & 0.2133 & 0.0820 \\ [6pt]
Our tree method & $N$ & & & & & \\
& 150 & 2.0000 & 1.1086 & 0.5155 & 0.2140 & 0.0825 \\
& 250 & 2.0000 & 1.1079 & 0.5190 & 0.2140 & 0.0822 \\
& 350 & 2.0000 & 1.1074 & 0.5193 & 0.2134 & 0.0828 \\[6pt]
Reference value & & 2.0000 & 1.1076 & 0.5200 & 0.2137 & 0.0820 \\
\hline
\end{tabular*}
\end{table}

\begin{table}
\def\arraystretch{0.95}
\caption{American put prices determined with our tree approach and
finite difference methods. Parameters: $K = 10$, $r = 0.1$, $\eta=
0.9$, $\kappa= 5.0$, $\theta= 0.16$, and
$\rho= 0.1$, $T=0.25$, $\sqrt{\nu_0} = \textbf{0.5}$}
\label{PDEput2}
\begin{tabular*}{\textwidth}{@{\extracolsep{\fill}}lcccccc@{}}
\hline
& & \multicolumn{5}{c@{}}{$\bolds{S_0}$} \\[-6pt]
& & \multicolumn{5}{c@{}}{\hrulefill} \\
\textbf{Method} & \textbf{Grid size} & \textbf{8} & \textbf{9} & \textbf{10} & \textbf{11} & \textbf{12} \\
\hline
PSOR & (40, 16, 8) & 2.0691 & 1.3139 & 0.7720 & 0.4293 & 0.2324 \\
& (60, 32, 66) & 2.0760 & 1.3292 & 0.7908 & 0.4442 & 0.2405 \\
& (120, 64, 130) & 2.0775 & 1.3320 & 0.7940 & 0.4467 & 0.2419 \\
& (240, 128, 258) & 2.0779 & 1.3329 & 0.7951 & 0.4476 & 0.2424 \\[6pt]
Componentwise & (40, 16, 8) & 2.0676 & 1.3094 & 0.7646 & 0.4232 & 0.2297
\\
splitting & (60, 32, 66) & 2.0758 & 1.3287 & 0.7900 & 0.4435 & 0.2401 \\
& (120, 64, 130) & 2.0774 & 1.3317 & 0.7936 & 0.4463 & 0.2417 \\
& (240, 128, 258) & 2.0780 & 1.3328 & 0.7949 & 0.4474 & 0.2423 \\ [6pt]
Transformation & (40, 16, 8) & 2.0691 & 1.3140 & 0.7721 & 0.4294 & 0.2325
\\
procedure & (60, 32, 66) & 2.0760 & 1.3291 & 0.7908 & 0.4442 & 0.2405 \\
& (120, 64, 130) & 2.0775 & 1.3319 & 0.7940 & 0.4467 & 0.2419 \\
& (240, 128, 258) & 2.0780 & 1.3329 & 0.7951 & 0.4476 & 0.2424 \\ [6pt]
Our tree method & $N$ & & & & & \\
& 150 & 2.0791 & 1.3362 & 0.7957 & 0.4495 & 0.2435 \\
& 250 & 2.0786 & 1.3338 & 0.7964 & 0.4501 & 0.2435 \\
& 350 & 2.0790 & 1.3339 & 0.7964 & 0.4485 & 0.2440 \\[6pt]
Reference value & & 2.0784 & 1.3336 & 0.7960 & 0.4483 & 0.2428 \\
\hline
\end{tabular*}
\end{table}

\subsection{Exotics}
Our numerical experimentation
confirms that
backward recursion yields quite fast and accurate results
for the two-dimensional problems
like European and American vanilla option pricing problems.
However,
our numerical
experimentation also reveals that the straightforward
application of the recursive method takes too long
on a personal computer when
another continuous variable is introduced
to price an exotic option. Hence, in order to substantially
speed up the computations, we
use our discrete equations
as a discretization scheme for our Monte Carlo (MC) simulation.
In other words, we carry out the MC simulation on the tree.

It is also important to note that our main concern in this section
is to show the pure application of our computation method. There are
many well-known
techniques in the literature which improve the speed and the accuracy
of tree and MC methods.
However, as in the backward recursion
we refrain from using any of these techniques.

Below we outline results for the geometric, arithmetic
Asian and for lookback options.

We start with the geometric Asian and let
\[
G_T = \exp \biggl( \frac{1}{T} \int_0^T
\ln(S_t)\,dt \biggr)
\]
be the geometric mean of $S_t$ over time $t$ during $[0,T]$. Then
the payoff of a fixed strike geometric Asian call is given by $\max(G_T-K,0)$.
Kim and Wee \cite{KW} provide semi-closed solutions for the price of geometric
Asian options under the Heston model. We compare our results with theirs.

\begin{table}
\def\arraystretch{0.94}
\caption{Comparison of our method and the semi-closed solution for
fixed-strike geometric Asian call options for: $S_0=100$, $\nu_0 =0.09$,
$r=0.05$, $\kappa=1.15$, $\theta= 0.348$, $\rho= -0.64$, $ \eta= 0.39$}
\label{geoAsian1}
\begin{tabular*}{\textwidth}{@{\extracolsep{\fill}}ld{3.0}d{2.4}d{2.4}d{2.4}d{2.4}d{2.2}d{2.2}d{2.2}@{}}
\hline
&  & \multicolumn{3}{c}{\textbf{MC on tree with} $\bolds{N=300}$} &
&\multicolumn{3}{c@{}}{\textbf{Difference \%}}\\[-6pt]
&  & \multicolumn{3}{c}{\hrulefill} &
&\multicolumn{3}{c@{}}{\hrulefill}\\
& & \multicolumn{3}{c}{\textbf{NumSim}} & & \multicolumn{3}{c@{}}{\textbf{NumSim}} \\[-6pt]
& & \multicolumn{3}{c}{\hrulefill} & & \multicolumn{3}{c@{}}{\hrulefill} \\
\multicolumn{1}{@{}l}{{$\bolds{T}$}}&
\multicolumn{1}{c}{$\bolds{K}$}& \multicolumn{1}{c}{$\bolds{10^5}$}& \multicolumn{1}{c}{$\bolds{5*10^5}$} & \multicolumn{1}{c}{$\bolds{10^6}$} &
\multicolumn{1}{c}{\multirow{3}{60pt}[12pt]{\centering\textbf{Semi-closed solution}}} &
\multicolumn{1}{c}{$\bolds{10^5}$} &\multicolumn{1}{c}{$\bolds{5*10^5}$} & \multicolumn{1}{c@{}}{$ \bolds{10^6}$}\\
\hline
0.2 & 90 & 10.6598 & 10.6551 & 10.6562 & 10.6571 & 0.02 & -0.02 & -0.01
\\
0.2 & 95 & 6.6006 & 6.5970 & 6.5888 & 6.5871 & 0.20 & 0.15 & 0.03 \\
0.2 & 100 & 3.4699 & 3.4564 & 3.4510 & 3.4478 & 0.64 & 0.25 & 0.09 \\
0.2 & 105 & 1.4697 & 1.4610 & 1.4611 & 1.4552 & 1.00 & 0.40 & 0.40 \\
0.2 & 110 & 0.4730 & 0.4742 & 0.4719 & 0.4724 & 0.14 & 0.38 & -0.10 \\
0.4 & 90 & 11.7310 & 11.7111 & 11.7077 & 11.7112 & 0.17 & 0.00 & -0.03
\\
0.4 & 95 & 8.0988 & 8.1067 & 8.0877 & 8.0894 & 0.12 & 0.21 & -0.02 \\
0.4 & 100 & 5.1480 & 5.1746 & 5.1641 & 5.1616 & -0.26 & 0.25 & 0.05 \\
0.4 & 105 & 3.0414 & 3.0060 & 3.0040 & 3.0018 & 1.32 & 0.14 & 0.07 \\
0.4 & 110 & 1.5555 & 1.5776 & 1.5679 & 1.5715 & -1.02 & 0.39 & -0.23 \\
0.5 & 90 & 12.2974 & 12.2495 & 12.2330 & 12.2329 & 0.53 & 0.14 & 0.00
\\
0.5 & 95 & 8.7711 & 8.7668 & 8.7753 & 8.7553 & 0.18 & 0.13 & 0.23 \\
0.5 & 100 & 5.9036 & 5.9151 & 5.9008 & 5.8971 & 0.11 & 0.31 & 0.06 \\
0.5 & 105 & 3.7150 & 3.7120 & 3.7165 & 3.7072 & 0.21 & 0.13 & 0.25 \\
0.5 & 110 & 2.1622 & 2.1692 & 2.1595 & 2.1589 & 0.15 & 0.48 & 0.03 \\
1 & 90 & 14.5646 & 14.6087 & 14.5937 & 14.5779 & -0.09 & 0.21 & 0.11 \\
1 & 95 & 11.6287 & 11.5518 & 11.5474 & 11.5551 & 0.64 & -0.03 & -0.07
\\
1 & 100 & 8.9708 & 8.9378 & 8.9530 & 8.9457 & 0.28 & -0.09 & 0.08 \\
1 & 105 & 6.8003 & 6.7392 & 6.7505 & 6.7559 & 0.66 & -0.25 & -0.08 \\
1 & 110 & 5.0161 & 4.9878 & 4.9704 & 4.9722 & 0.88 & 0.31 & -0.04 \\
1.5 & 90 & 16.3889 & 16.4588 & 16.5200 & 16.5030 & -0.69 & -0.27 & 0.10
\\
1.5 & 95 & 13.7324 & 13.7764 & 13.7690 & 13.7625 & -0.22 & 0.10 & 0.05
\\
1.5 & 100 & 11.3599 & 11.3247 & 11.3304 & 11.3374 & 0.20 & -0.11 &
-0.06 \\
1.5 & 105 & 9.2487 & 9.2187 & 9.2076 & 9.2245 & 0.26 & -0.06 & -0.18 \\
1.5 & 110 & 7.4342 & 7.3959 & 7.4019 & 7.4122 & 0.30 & -0.22 & -0.14 \\
2 & 90 & 18.0757 & 18.1112 & 18.0816 & 18.0914 & -0.09 & 0.11 & -0.05
\\
2 & 95 & 15.6133 & 15.6021 & 15.5211 & 15.5640 & 0.32 & 0.24 & -0.28 \\
2 & 100 & 13.3624 & 13.3245 & 13.2833 & 13.2933 & 0.52 & 0.24 & -0.08
\\
2 & 105 & 11.2855 & 11.2862 & 11.2627 & 11.2728 & 0.11 & 0.12 & -0.09
\\
2 & 110 & 9.4243 & 9.4840 & 9.4901 & 9.4921 & -0.71 & -0.09 & -0.02 \\
3 & 90 & 20.6523 & 20.4276 & 20.5149 & 20.5102 & 0.69 & -0.40 & 0.02 \\
3 & 95 & 18.3985 & 18.2361 & 18.2884 & 18.3060 & 0.51 & -0.38 & -0.10
\\
3 & 100 & 16.2151 & 16.2555 & 16.2609 & 16.2895 & -0.46 & -0.21 & -0.18
\\
3 & 105 & 14.5000 & 14.4330 & 14.4046 & 14.4531 & 0.32 & -0.14 & -0.34
\\
3 & 110 & 12.6065 & 12.8177 & 12.7982 & 12.7882 & -1.42 & 0.23 & 0.08
\\
\hline
\end{tabular*}  \vspace*{-3pt}
\end{table}

\begin{table}
\def\arraystretch{0.95}
\caption{Confidence intervals for fixed-strike geometric Asian call
options for: $S_0=100$, $\nu_0 =0.09$,
$r=0.05$, $\kappa=1.15$, $\theta= 0.348$, $\rho= -0.64$, $ \eta=
0.39$}\vspace*{-3pt}
\label{geoAsian2}
\begin{tabular*}{\textwidth}{@{\extracolsep{\fill}}lcc@{}}\hline
\multicolumn{3}{c}{\textbf{Confidence intervals 95\%}} \\
$\bolds{\mathrm{NumSim} = 10^5}$ & $ \bolds{\mathrm{NumSim} = 5*10^5} $ & \multicolumn{1}{c@{}}{$\bolds{\mathrm{NumSim} = 10^6}$} \\
\hline
$ ( 10.6135, 10.7060 ) $ & $ ( 10.6345, 10.6758 ) $ & $ ( 10.6416,
10.6708 ) $ \\
$ ( 6.5609, 6.6402 ) $ & $ ( 6.5793, 6.6147 ) $ & $ ( 6.5763, 6.6014
) $ \\
$ ( 3.4397, 3.5001 ) $ & $ ( 3.4429, 3.4699 ) $ & $ ( 3.4415, 3.4605
) $ \\
$ ( 1.4501, 1.4894 ) $ & $ ( 1.4522, 1.4698 ) $ & $ ( 1.4548, 1.4673
) $ \\
$ ( 0.4623, 0.4838 ) $ & $ ( 0.4694, 0.4790 ) $ & $ ( 0.4685, 0.4753
) $ \\
$ ( 11.6678, 11.7941 ) $ & $ ( 11.6829, 11.7394 ) $ & $ ( 11.6877,
11.7277 ) $ \\
$ ( 8.0438, 8.1538 ) $ & $ ( 8.0820, 8.1313 ) $ & $ ( 8.0703, 8.1051
) $ \\
$ ( 5.1027, 5.1932 ) $ & $ ( 5.1543, 5.1948 ) $ & $ ( 5.1498, 5.1784
) $ \\
$ ( 3.0065, 3.0764 ) $ & $ ( 2.9904, 3.0216 ) $ & $ ( 2.9930, 3.0150
) $ \\
$ ( 1.5308, 1.5803 ) $ & $ ( 1.5665, 1.5887 ) $ & $ ( 1.5601, 1.5758
) $ \\
$ ( 12.2270, 12.3679 ) $ & $ ( 12.2181, 12.2808 ) $ & $ ( 12.2108,
12.2552 ) $ \\
$ ( 8.7094, 8.8328 ) $ & $ ( 8.7391, 8.7944 ) $ & $ ( 8.7557, 8.7949
) $ \\
$ ( 5.8516, 5.9556 ) $ & $ ( 5.8919, 5.9384 ) $ & $ ( 5.8843, 5.9172
) $ \\
$ ( 3.6735, 3.7566 ) $ & $ ( 3.6934, 3.7306 ) $ & $ ( 3.7034, 3.7297
) $ \\
$ ( 2.1305, 2.1938 ) $ & $ ( 2.1551, 2.1833 ) $ & $ ( 2.1495, 2.1694
) $ \\
$ ( 14.4642, 14.6650 ) $ & $ ( 14.5638, 14.6536 ) $ & $ ( 14.5619,
14.6255 ) $ \\
$ ( 11.5367, 11.7208 ) $ & $ ( 11.5109, 11.5927 ) $ & $ ( 11.5186,
11.5763 ) $ \\
$ ( 8.8888, 9.0528 ) $ & $ ( 8.9013, 8.9744 ) $ & $ ( 8.9272, 8.9789
) $ \\
$ ( 6.7282, 6.8724 ) $ & $ ( 6.7072, 6.7713 ) $ & $ ( 6.7278, 6.7732
) $ \\
$ ( 4.9538, 5.0784 ) $ & $ ( 4.9601, 5.0154 ) $ & $ ( 4.9508, 4.9899
) $ \\
$ ( 16.2635, 16.5144 ) $ & $ ( 16.4023, 16.5152 ) $ & $ ( 16.4800,
16.5599 ) $ \\
$ ( 13.6150, 13.8498 ) $ & $ ( 13.7239, 13.8289 ) $ & $ ( 13.7319,
13.8061 ) $ \\
$ ( 11.2523, 11.4676 ) $ & $ ( 11.2765, 11.3729 ) $ & $ ( 11.2963,
11.3645 ) $ \\
$ ( 9.1503, 9.3471 ) $ & $ ( 9.1749, 9.2626 ) $ & $ ( 9.1766, 9.2387
) $ \\
$ ( 7.3457, 7.5226 ) $ & $ ( 7.3563, 7.4355 ) $ & $ ( 7.3739, 7.4299
) $ \\
$ ( 17.9261, 18.2253 ) $ & $ ( 18.0442, 18.1782 ) $ & $ ( 18.0342,
18.1289 ) $ \\
$ ( 15.4721, 15.7544 ) $ & $ ( 15.5392, 15.6651 ) $ & $ ( 15.4767,
15.5654 ) $ \\
$ ( 13.2303, 13.4945 ) $ & $ ( 13.2656, 13.3835 ) $ & $ ( 13.2416,
13.3249 ) $ \\
$ ( 11.1626, 11.4084 ) $ & $ ( 11.2315, 11.3409 ) $ & $ ( 11.2240,
11.3013 ) $ \\
$ ( 9.3113, 9.5373 ) $ & $ ( 9.4334, 9.5345 ) $ & $ ( 9.4544, 9.5258
) $ \\
$ ( 20.4610, 20.8436 ) $ & $ ( 20.3429, 20.5123 ) $ & $ ( 20.4547,
20.5750 ) $ \\
$ ( 18.2156, 18.5814 ) $ & $ ( 18.1549, 18.3174 ) $ & $ ( 18.2310,
18.3459 ) $ \\
$ ( 16.0417, 16.3885 ) $ & $ ( 16.1781, 16.3329 ) $ & $ ( 16.2062,
16.3157 ) $ \\
$ ( 14.3343, 14.6656 ) $ & $ ( 14.3594, 14.5067 ) $ & $ ( 14.3525,
14.4566 ) $ \\
$ ( 12.4514, 12.7617 ) $ & $ ( 12.7476, 12.8878 ) $ & $ ( 12.7489,
12.8476 ) $ \\
\hline
\end{tabular*}     \vspace*{-5pt}
\end{table}

Table~\ref{geoAsian1} displays a comparison between prices from the
semi-closed solution and
those from our MC simulation\vadjust{\goodbreak} on tree with $N =300$ and number of simulations
$(\mathrm{NumSim})= 10^5, 5*10^5, 10^ 6$.
As benchmark prices, we use the values given in Table~5 from
\cite{KW} for the parameter values: $S_0=100$, $\nu_0 =0.09$, $r=0.05$,
$\kappa=1.15$, $\theta= 0.348$, $\rho= -0.64$, $ \eta= 0.39$. As it
is clear
from the table, our numerical scheme provides a very good approximation
for the
analytical prices. For $\mathrm{NumSim} = 10^6$, we get the three largest
percentage errors
as $( 0.40 \%, 0.34 \%, 0.28 \% )$ and the average percentage error is
$ 0.11 \%$.
Table~\ref{geoAsian2} shows the $95 \%$ confidence intervals for the
prices computed for
different numbers of simulations.\vadjust{\vspace*{-3pt}\goodbreak}

Table~\ref{AAsian} includes our results for arithmetic Asian options
under the Heston model.
We carry out the simulations as in the same way described previously.
Let
\[
A_T = \exp \biggl( \frac{1}{T} \int_0^T
S_t \,dt \biggr)
\]
be the arithmetic average of $S_t$ over time $t$ during $[0,T]$. Then
the payoff of a fixed strike arithmetic Asian call is given by $\max(A_T-K,0)$.
Pages and Printems~\cite{PP} use the functional quantization based
quadrature formula
to price vanilla calls and Asian calls in the Heston model. The numbers computed
from MC method, Romberg log-extrapolation and $K$-interpolation of
Romberg and
their standard deviations in the parenthesis are tabulated for
comparison; see Table~4 in \cite{PP} for a more detailed explanation of
the results.
We test our model using the numbers reported in their paper. As one can observe
from Table~\ref{AAsian}, our prices together with the confidence
intervals are in accordance with the only reference values for arithmetic
Asian options under the Heston dynamics which can be found in the literature.

%
\begin{table}
\tabcolsep=4pt
\caption{Comparison of our method and the functional quantization
method by Pages and Printems \cite{PP} for arithmetic Asian options.
Parameters: $S_0=50$, $\nu_0 =0.01$, $r=0.05$, $\kappa=2$, $\theta=
0.01$, $\rho= 0.5$, $ \eta= 0.1$}
\label{AAsian}
\begin{tabular*}{\textwidth}{@{\extracolsep{\fill}}ld{1.3}cccd{1.3}ccc@{}}
\hline
& & & & & &&\multicolumn{2}{c@{}}{\textbf{Our method}}\\[-6pt]
& & & & & &&\multicolumn{2}{c@{}}{\hrulefill}\\
& & & & & &&\multicolumn{2}{c@{}}{$\bolds{N=300,}\ \mathbf{NumSim \bolds{=}10^6}$} \\[-6pt]
& & & & & &&\multicolumn{2}{c@{}}{\hrulefill}\\
\multicolumn{1}{@{}l}{{$\bolds{K}$}} &
\multicolumn{1}{c}{{$\bolds{10^8}$\textbf{-MC}}} &
\multicolumn{1}{c}{\multirow{2}{50pt}[10pt]{\centering\textbf{Crude MC reference}}} &
\multicolumn{2}{c}{\multirow{2}{50pt}[10pt]{\centering\textbf{Romberg on crude FQ}}} &
\multicolumn{2}{c}{\multirow{2}{50pt}[10pt]{\centering\textbf{$\bolds{K}$-interpol. of Romberg}}} & \multicolumn{1}{c}{\textbf{Price}} &
\multicolumn{1}{c@{}}{\textbf{Conf. int.}} \\
\hline
44 & 6.92 & $ (0.08 \%) $ & 6.92 & $ (0.01 \%) $ & 6.92 & $ (0.01 \%) $
& 6.9196 & (6.9139, 6.9252) \\
45 & 5.97 & $ (0.10 \%) $ & 5.97 & $ (0.04 \%) $ & 5.97 & $ (0.02 \%) $
& 5.9768 & (5.9712, 5.9825) \\
46 & 5.03 & $ (0.11 \%) $ & 5.03 & $ (0.05 \%) $ & 5.03 & $ (0.02 \%) $
& 5.0334 & (5.0278, 5.0390) \\
47 & 4.11 & $ (0.14 \%) $ & 4.12 & $ (0.09 \%) $ & 4.11 & $ (0.04 \%) $
& 4.1117 & (4.1062, 4.1172) \\
48 & 3.245 & $ (0.16 \%) $ & 3.25 & $ (0.17 \%) $ & 3.24 & $ (0.05 \%)
$ & 3.2506 & (3.2453, 3.2559) \\
49 & 2.46 & $ (0.20 \%) $ & 2.47 & $ (0.32 \%) $ & 2.46 & $ (0.04 \%) $
& 2.4673 & (2.4624, 2.4723) \\
50 & 1.79 & $ (0.26 \%) $ & 1.80 & $ (0.63 \%) $ & 1.79 & $ (0.03 \%) $
& 1.7926 & (1.7882, 1.7970) \\
51 & 1.25 & $ (0.31 \%) $ & 1.26 & $ (1.16 \%) $ & 1.25 & $ (0.17 \%) $
& 1.2541 & (1.2503, 1.2580) \\
52 & 0.84 & $ (0.39 \%) $ & 0.85 & $ (2.06 \%) $ & 0.84 & $ (0.37 \%) $
& 0.8430 & (0.8398, 0.8463) \\
53 & 0.54 & $ (0.50 \%) $ & 0.56 & $ (3.73 \%) $ & 0.545 & $ (0.78 \%)
$ & 0.5502 & (0.5475, 0.5529) \\
54 & 0.34 & $ (0.63 \%) $ & 0.36 & $ (6.58 \%) $ & 0.34 & $ (1.37 \%) $
& 0.3485 & (0.3464, 0.3506) \\
55 & 0.21 & $ (0.81 \%) $ & 0.23 & $ (11.53 \%) $ & 0.21 & $ (2.15 \%)
$ & 0.2159 & (0.2142, 0.2176) \\
56 & 0.125 & $ (1.04 \%) $ & 0.15 & $ (19.96 \%) $ & 0.125 & $ (2.84 \%
) $ & 0.1317 & (0.1303, 0.1330) \\
\hline
\end{tabular*}
\end{table}
%
%
\begin{table}
\tabcolsep=0pt
\def\arraystretch{0.95}
\caption{Comparison of our method and Euler simulation for lookback
call option with fixed strike. Parameters: $S_0=100$, $\nu_0 =0.16$,
$r=0.05$, $\kappa=3$, $\theta= 0.04$, $\rho= -0.7$, $ \eta= 0.1$}
\label{lookback}
\begin{tabular*}{\textwidth}{@{\extracolsep{\fill}}ld{3.0}d{2.4}cd{2.4}cc@{}}
\hline
& & \multicolumn{2}{c}{\textbf{Euler simulation}} & \multicolumn{2}{c}{\textbf{Our method}}&\\[-6pt]
& & \multicolumn{2}{c}{\hrulefill} & \multicolumn{2}{c}{\hrulefill}&\\
& & \multicolumn{2}{c}{$\bolds{N=3000}$\textbf{,} $\mathbf{NumSim}\bolds{ = 10^5}$} &
\multicolumn{2}{c}{$\bolds{N=3000}$\textbf{,} $\mathbf{NumSim}\bolds{ = 10^5}$} & \\[-6pt]
& & \multicolumn{2}{c}{\hrulefill} & \multicolumn{2}{c}{\hrulefill}&\\
\multicolumn{1}{@{}l}{{$\bolds{T}$}}&
\multicolumn{1}{c}{{$\bolds{K}$}}& \multicolumn{1}{c}{\textbf{Price}} & \multicolumn{1}{c}{\textbf{Confidence interval}} &
\multicolumn{1}{c}{\textbf{Price}} &
\multicolumn{1}{c}{\textbf{Confidence interval}} & \multicolumn{1}{c@{}}{{\textbf{Difference} $\bolds{\%}$}}\\
\hline
0.2 & 90 & 23.4527 & (23.3844, 23.5210) & 23.4679 & (23.3996,
23.5362) & 0.06 \\
0.2 & 95 & 18.5511 & (18.4827, 18.6196) & 18.5459 & (18.4776,
18.6142) & 0.03 \\
0.2 & 100 & 13.5145 & (13.4464, 13.5825) & 13.6562 & (13.5878,
13.7246) & 1.05 \\
0.2 & 105 & 9.2629 & (9.1987, 9.3272) & 9.2620 & (9.1978, 9.3262)
& 0.01 \\
0.2 & 110 & 6.0746 & (6.0185, 6.1306) & 6.0899 & (6.0340, 6.1457)
& 0.25 \\
0.4 & 90 & 27.7252 & (27.6333, 27.8172) & 27.7378 & (27.6461,
27.8296) & 0.05 \\
0.4 & 95 & 22.7931 & (22.7015, 22.8846) & 22.7784 & (22.6869,
22.8698) & 0.06 \\
0.4 & 100 & 17.8937 & (17.8017, 17.9857) & 17.9052 & (17.8136,
17.9969) & 0.06 \\
0.4 & 105 & 13.5301 & (13.4415, 13.6187) & 13.6541 & (13.5649,
13.7434) & 0.92 \\
0.4 & 110 & 10.0038 & (9.9224, 10.0852) & 10.0978 & (10.0160,
10.1796) & 0.94 \\
0.5 & 90 & 29.1737 & (29.0738, 29.2735) & 29.2407 & (29.1405,
29.3409) & 0.23 \\
0.5 & 95 & 24.2728 & (24.1733, 24.3722) & 24.3094 & (24.2095,
24.4093) & 0.15 \\
0.5 & 100 & 19.4547 & (19.3542, 19.5552) & 19.5036 & (19.4033,
19.6038) & 0.25 \\
0.5 & 105 & 15.1074 & (15.0099, 15.2049) & 15.0772 & (14.9801,
15.1742) & 0.20 \\
0.5 & 110 & 11.4637 & (11.3730, 11.5544) & 11.4401 & (11.3498,
11.5305) & 0.21 \\
1 & 90 & 34.1211 & (33.9910, 34.2511) & 34.1944 & (34.0646,
34.3242) & 0.21 \\
1 & 95 & 29.4579 & (29.3273, 29.5886) & 29.4015 & (29.2720,
29.5311) & 0.19 \\
1 & 100 & 24.6878 & (24.5573, 24.8184) & 24.7163 & (24.5855,
24.8470) & 0.12 \\
1 & 105 & 20.1960 & (20.0686, 20.3234) & 20.3721 & (20.2443,
20.4999) & 0.87 \\
1 & 110 & 16.5429 & (16.4206, 16.6652) & 16.4579 & (16.3367,
16.5791) & 0.51 \\
1.5 & 90 & 37.6113 & (37.4587, 37.7640) & 37.8563 & (37.7035,
38.0091) & 0.65 \\
1.5 & 95 & 33.2861 & (33.1314, 33.4408) & 33.0959 & (32.9428,
33.2491) & 0.57 \\
1.5 & 100 & 28.5915 & (28.4380, 28.7451) & 28.3913 & (28.2386,
28.5440) & 0.70 \\
1.5 & 105 & 24.2427 & (24.0913, 24.3941) & 24.1616 & (24.0107,
24.3124) & 0.33 \\
1.5 & 110 & 20.4593 & (20.3131, 20.6054) & 20.4385 & (20.2919,
20.5850) & 0.10 \\
2 & 90 & 41.0722 & (40.8963, 41.2481) & 41.0605 & (40.8861,
41.2350) & 0.03 \\
2 & 95 & 36.6204 & (36.4454, 36.7953) & 36.5932 & (36.4185,
36.7680) & 0.07 \\
2 & 100 & 31.9362 & (31.7612, 32.1112) & 32.0618 & (31.8874,
32.2361) & 0.39 \\
2 & 105 & 27.8954 & (27.7220, 28.0688) & 27.7302 & (27.5578,
27.9026) & 0.59 \\
2 & 110 & 24.0406 & (23.8719, 24.2093) & 23.8907 & (23.7223,
24.0591) & 0.62 \\
3 & 90 & 47.0043 & (46.7881, 47.2205) & 47.0854 & (46.8698,
47.3010) & 0.17 \\
3 & 95 & 42.6606 & (42.4453, 42.8759) & 42.5750 & (42.3599,
42.7901) & 0.20 \\
3 & 100 & 38.6746 & (38.4588, 38.8903) & 38.3630 & (38.1469,
38.5790) & 0.81 \\
3 & 105 & 34.5038 & (34.2898, 34.7177) & 34.2793 & (34.0657,
34.4929) & 0.65 \\
3 & 110 & 30.7339 & (30.5229, 30.9449) & 30.4407 & (30.2312,
30.6502) & 0.95 \\
\hline
\end{tabular*}
\end{table}

It is clear that when we price a lookback option using backward recursion,
we also need another continuous variable holding the running
max or min. But in this case, we can constrain this variable to
take values on a tree as well. However, it still remains more
efficient to apply our MC method on the tree.
Table~\ref{lookback} presents numerical results obtained by the
standard MC method
and our numerical method for fixed strike lookback call options.
As comparison we used simple Monte Carlo simulations
based on a Euler method. The table
contains prices for $N = 3000$ and $\mathrm{NumSim} = 10^5$. As one can see from the
last column, the numbers obtained from our numerical method differ only
slightly from the prices computed by
the Euler MC method.

In terms of the theoretical complexity, we require $n^3$ many
computations for
$n$ many time steps in the difference equations case. This is similar
to that of PDE approach.
More precisely, Table~\ref{timetable} provides average running times
for the options in Tables~\ref{HestonEput}--\ref{lookback}.
The computer used is a standard laptop with an Intel Core i7 M620@2.67
GHz CPU and a 4 GB memory.
The algorithm was implemented in MATLAB.

%
\begin{table}
\caption{Average running times for options in
Tables~\protect\ref{HestonEput}--\protect\ref{lookback}}
\label{timetable}
\begin{tabular*}{\textwidth}{@{\extracolsep{\fill}}lcd{2.2}@{}}
\hline
& & \multicolumn{1}{c@{}}{\textbf{Time in seconds}} \\
\hline
European put and call & $N=200$ & 5.71 \\
\quad (Tables~\ref{HestonEput} and \ref{HestonEcall}) & $N=350$ & 30.37 \\
 & $N=500$ & 89.27 \\[3pt]
American put & {$N=250$} &13.97 \\
\quad (Table~\ref{NBput}) & & \\[3pt]
American put & $N=150$ & 3.15 \\
\quad (Tables~\ref{PDEput1} and \ref{PDEput2}) & $N=250$ & 14.66 \\
  & $N=350$ & 40.50 \\[3pt]
Geometric Asian & $\mathrm{NumSim}=10^5$ & 8.17 \\
\quad (Tables~\ref{geoAsian1} and \ref{geoAsian2})  & $\mathrm{NumSim}=5*10^5$ & 40.91 \\
& $\mathrm{NumSim}=10^6$ & 81.79 \\[3pt]
Arithmetic Asian& $N=300$ &98.65 \\
\quad (Table~\ref{AAsian}) & {$\mathrm{NumSim}=10^6$}& \\[3pt]
{Lookback} & $N=3000$ & 94.54 \\
\quad (Table~\ref{lookback}) & {$\mathrm{NumSim}=10^5$}& \\
\hline
\end{tabular*}
\end{table}

\section{Concluding remarks}

In this paper, we have developed
a recombining tree approximation
of the Heston model. Our approach
is very general and applies to all
stochastic volatility models with a factor equation.
Low-dimensional European
and American option equations
can be solved by a straightforward
backward recursion. We have done extensive numerical experimentation
with the resulting pricing equations. These results,
reported in the previous section,
confirm the efficiency of the method.
%
%

%

%




\printaddresses

\end{document}